\documentclass[journal]{IEEEtran}

\usepackage{amssymb}
\usepackage{amsmath}
\usepackage{cite}
\usepackage{url}
\usepackage{empheq}
\usepackage{xcolor}
\usepackage{graphicx}
\usepackage{subfigure}
\usepackage{enumitem}
\usepackage{fancyhdr}
\usepackage{mdwmath}	
\usepackage{mdwtab}
\usepackage{caption}
\usepackage{amsthm}
\usepackage{algorithm}
\usepackage{algorithmic}

\newtheorem{lemma}{Lemma}
\newtheorem{remark}{Remark}
\newtheorem{theorem}{Theorem}
\newtheorem{corollary}{Corollary}
\newtheorem{assumption}{Assumption}
\newtheorem{definition}{Definition}
\newtheorem{proposition}{Proposition}

\newcommand{\fref}[1]{Fig.~\ref{#1}}

\newcommand*{\QEDA}{\null\nobreak\hfill\ensuremath{\blacksquare}}

\begin{document}
\title{Trajectory Design and User Scheduling for Pinching-Antenna Systems}
\author{Kaidi~Wang,~\IEEEmembership{Member,~IEEE,}
Daniel~K.~C.~So,~\IEEEmembership{Senior~Member,~IEEE,}
and~Zhiguo~Ding,~\IEEEmembership{Fellow,~IEEE}
\thanks{Kaidi~Wang and Daniel~K.~C.~So are with the Department of Electrical and Electronic Engineering, the University of Manchester, Manchester, M1 9BB, UK (e-mail: kaidi.wang@ieee.org; d.so@manchester.ac.uk).}
\thanks{Zhiguo~Ding is with the School of Electrical and Electronic Engineering (EEE), Nanyang Technological University, Singapore 639798 (e-mail: zhiguo.ding@ntu.edu.sg).}}
\maketitle
%%%%%%%%%%%%%%%%%%%%%%%%%%%%%%%%%%%%%%%%%%%%%%%%%
%%%%%%%%%%%%%%%%%%%%%%%%%%%%%%%%%%%%%%%%%%%%%%%%%
\begin{abstract}
Finite antenna movement speed introduces repositioning delays that can limit the performance gains of pinching-antenna systems. In this paper, a trajectory based scheme is proposed for a time division multiple access (TDMA) based pinching-antenna system, allowing transmission during both movement and stationary operation. A discrete-time model is developed to characterize periodic antenna trajectories and antenna-user service association under practical movement constraints. For a single-waveguide scenario, the expected average achievable rate is derived. The rate loss relative to instantaneous positioning and the rate improvement over a fixed antenna are further analyzed, with an upper bound on the former and a lower bound on the latter established. To improve fairness among users, a minimum average rate maximization problem is formulated through joint trajectory design and user scheduling. To address the coupling between these variables, an alternating optimization (AO) framework is adopted. Specifically, a successive convex approximation (SCA) based algorithm is developed using concave rate lower bounds, while a game based scheduling algorithm employs an ordered user-rate vector as the common utility. The convergence of the SCA objective value and the finite termination and stability of the scheduling algorithm are established. Simulation results demonstrate the gains achieved by the joint design and show that the proposed system can approach instantaneous-positioning performance under finite antenna movement speed.
\end{abstract}
%%%%%%%%%%%%%%%%%%%%%%%%%%%%%%%%%%%%%%%%%%%%%%%%% 
%%%%%%%%%%%%%%%%%%%%%%%%%%%%%%%%%%%%%%%%%%%%%%%%%
\section{Introduction}
The increasing demand for higher data rates and wider coverage has motivated research into flexible antenna technologies, such as reconfigurable intelligent surfaces (RISs), fluid antennas, and movable antennas, which provide greater spatial adaptability than conventional fixed antennas. Specifically, RISs redirect incident signals by adjusting the responses of reflecting elements \cite{wu2019irs}. Fluid antennas enable position reconfiguration through fluidic or switchable structures \cite{wong2020fluid}, while movable antennas mechanically reposition antenna elements to exploit spatial channel variations \cite{zhu2024movable}. Although these technologies provide effective spatial control, their performance gains remain subject to certain limitations. RIS assisted transmission involves both transmitter-RIS and RIS-receiver paths, leading to cascaded propagation losses \cite{pan2021ris}. In typical compact implementations, fluid and movable antennas adjust their positions over regions of a few wavelengths, primarily exploiting small-scale channel variations with limited changes in propagation distance \cite{wu2024fluid, ma2024movable}. Therefore, further extending the range of radiating positions while limiting propagation loss remains an important direction for antenna design.

In this context, pinching-antenna systems provide a promising solution that combines flexible positioning over extended regions with reduced propagation loss \cite{suzuki2022pinching, ding2024pin}. In this architecture, small dielectric particles, referred to as pinches, are placed along dielectric waveguides to couple guided signals into free space, creating localized radiation points \cite{kaidi2025pin}. By employing low-loss waveguides to deliver signals to radiation points closer to the users, these systems can shorten the free-space propagation distance, thereby significantly improving the channel gains \cite{yang2025pinching}. The flexibility in antenna positioning also facilitates the selection of unblocked propagation paths, allowing line-of-sight (LoS) links to be established when suitable radiation locations are available \cite{xu2025generalized}. Moreover, the positions of multiple pinching antennas can be jointly configured to align the signal phases at the intended receiver, enabling coherent combining through pinching beamforming \cite{liu2025pinching}. In addition, the simple structure of dielectric waveguides and compact pinching elements offers the potential for low-cost implementation \cite{lu2026pin}.
%%%%%%%%%%%%%%%%%%%%%%%%%%%%%%%%%%%%%%%%%%%%%%%%%
\subsection{Related Works}
Motivated by these advantages, existing studies have investigated \emph{discrete activation} and \emph{continuous positioning} in pinching-antenna systems. Discrete activation selects the locations of active pinching antennas from a finite set of candidates, corresponding either to pre-installed antennas at fixed locations or to candidate positions obtained by discretizing the continuous waveguide domain \cite{kaidi2025pin2, sun2026pin, liu2026pin, xu2026environment, kaidi2026pin3}. Specifically, a multi-waveguide non-orthogonal multiple access (NOMA) system with pre-installed pinching antennas was explored in \cite{kaidi2025pin2}, where waveguide assignment, antenna activation, decoding order design, and power allocation were jointly considered. For downlink and uplink multi-user systems, an element-wise optimization framework was developed in \cite{sun2026pin}, where antenna positions were sequentially selected from discretized candidate locations to maximize the sum rate. Discrete activation was also investigated for wireless powered mobile edge computing (MEC) in \cite{liu2026pin}. The authors compared time division multiple access (TDMA) with NOMA by jointly optimizing antenna activation and resource allocation to maximize the total number of computed bits. At the network level, discrete activation was optimized under given blockage layouts to maximize coverage or the minimum average signal-to-noise ratio (SNR) across the service region \cite{xu2026environment}. Beyond dielectric waveguide implementations, a leaky coaxial cable (LCX) based system was studied in \cite{kaidi2026pin3}, where binary activation of fixed radiation slots was extended to continuous aperture adjustment. Both static and user-specific aperture configurations were optimized under TDMA to improve the sum rate.

In contrast, continuous positioning allows pinching antennas to be positioned at arbitrary locations along the waveguides, providing greater spatial flexibility and a larger design space for antenna placement optimization \cite{chen2026pin, zhang2026pin, gan2026pass, mao2026outage, chen2026pinching}. The trade-off between waveguide attenuation and free-space path loss was investigated in \cite{chen2026pin}, where suboptimal positioning rules for a single pinching antenna serving multiple users were derived by maximizing a lower bound on the sum rate. A directional pinching-antenna model incorporating waveguide attenuation and stochastic LoS blockage was developed in \cite{zhang2026pin}. Single-antenna positions and orientations were analytically characterized, while antenna-user assignment and beamforming were optimized for multi-user transmission. For NOMA based downlink communications, the authors of \cite{gan2026pass} jointly optimized transmit beamforming, antenna positions, and power allocation to minimize the transmit power subject to user rate requirements. To support reliable multi-user transmission, an environment division multiple access (EDMA) framework was studied in \cite{mao2026outage}, where antenna positions, power allocation, and information rates were jointly optimized subject to statistical rate outage constraints under probabilistic LoS blockage and random non-LoS scattering. At the multi-cell level, the authors of \cite{chen2026pinching} jointly optimized precoding, antenna positions, and power allocation across pinching antennas to maximize the weighted sum rate in the presence of intercell interference.
%%%%%%%%%%%%%%%%%%%%%%%%%%%%%%%%%%%%%%%%%%%%%%%%%
\subsection{Motivation and Contributions}
To explore the performance limits of pinching-antenna systems, continuous positioning has been extensively investigated, typically with antenna repositioning time neglected \cite{ding2024pin}. In practice, however, finite antenna movement speed introduces repositioning delays that reduce the effective transmission time and average achievable rate when transmission is suspended during movement \cite{kaidi2026fl, xu2026pin2}. To address this issue, an antenna roaming scheme was developed in \cite{kaidi2026roaming}, allowing transmission during antenna movement. Building on this operating mode, this paper proposes a one-dimensional trajectory based pinching-antenna transmission scheme, in which antenna movement and stationary operation are planned over the service period to serve users under finite movement speed. This design provides greater flexibility in antenna reconfiguration under practical movement constraints, but also introduces additional optimization challenges. In particular, each antenna position needs to be jointly planned with those at adjacent time instants to ensure that the required displacement can be completed within the available time. Meanwhile, user scheduling determines the serving antenna and service intervals for each user, and therefore needs to be coordinated with antenna trajectories to serve users at favorable antenna positions. These interdependencies motivate the joint design of antenna trajectories and user scheduling.

The main contributions are summarized as follows:
\begin{itemize}[leftmargin=*]
\item A practical trajectory based pinching-antenna system is proposed, with antenna movement modeled in discrete time over the service period. This model incorporates position, movement speed, and periodicity constraints to ensure feasible trajectories, while antenna-user association specifies the active antenna and scheduled user in each time slot. The benefits of repositioning inactive antennas in parallel for multi-waveguide scenarios are also indicated.
\item To characterize the performance of the proposed scheme, the expected average achievable rate is derived for a single-waveguide scenario. Based on this result, the rate loss relative to instantaneous positioning and the rate improvement over a fixed antenna are quantified, with an upper bound on the former and a lower bound on the latter further established. These analytical results reveal the effects of key system parameters on the achievable performance.
\item Considering fairness among users, a minimum average rate maximization problem is formulated through joint trajectory design and user scheduling. This formulation captures the coupling through the channel conditions and allocated service duration that determine each user's average rate. To facilitate the solution, the problem is decomposed into two subproblems, which are solved iteratively within an alternating optimization (AO) framework.
\item For trajectory design, a successive convex approximation (SCA) based algorithm is developed using tight concave rate lower bounds, with objective convergence established. For user scheduling, a game based algorithm employs an ordered user-rate vector as the common utility and updates configurations through pair construction and pair exchange. Its complexity, finite termination, and Nash and exchange stability under fixed trajectories are analyzed.
\end{itemize}
Simulation results are presented to evaluate the proposed trajectory based scheme and the developed algorithms. The results demonstrate that joint trajectory design and user scheduling improve the minimum average achievable rate, enabling the system to approach the instantaneous-positioning benchmark under finite antenna movement speed. The effects of key system parameters are also investigated, indicating the benefits of coordinating antenna movement with user service.
%%%%%%%%%%%%%%%%%%%%%%%%%%%%%%%%%%%%%%%%%%%%%%%%%
%%%%%%%%%%%%%%%%%%%%%%%%%%%%%%%%%%%%%%%%%%%%%%%%%
\section{System Model}
Consider a downlink pinching-antenna system with $K$ dielectric waveguides and $N$ users, where each waveguide is equipped with a single pinching antenna. The collections of all waveguides and users are denoted by $\mathcal{K}=\{1,2,\cdots,K\}$ and $\mathcal{N}=\{1,2,\cdots,N\}$, respectively. The system is deployed over a rectangular service region of size $D_x\times D_y$, with the center designated as the origin of the 3D Cartesian coordinate system. The waveguides are deployed parallel to the $x$-axis at a height of $d$, where waveguide $k$ is located at the fixed $y$-coordinate $y_k$, with the feed point positioned at the left end, i.e., $\boldsymbol{\psi}_k^\mathrm{feed}=(-D_x/2,y_k,d)$. The pinching antenna deployed on waveguide $k$, also referred to as pinching antenna $k$, can move along the $x$-axis. The users are randomly distributed on the ground within the service region, with the position of user $n$ denoted by $\boldsymbol{\psi}_n=(x_n,y_n,0)$.
%%%%%%%%%%%%%%%%%%%%%%%%%%%%%%%%%%%%%%%%%%%%%%%%%
\subsection{Trajectory Model}
\begin{figure}[!t]
\centering\includegraphics[width=85mm]{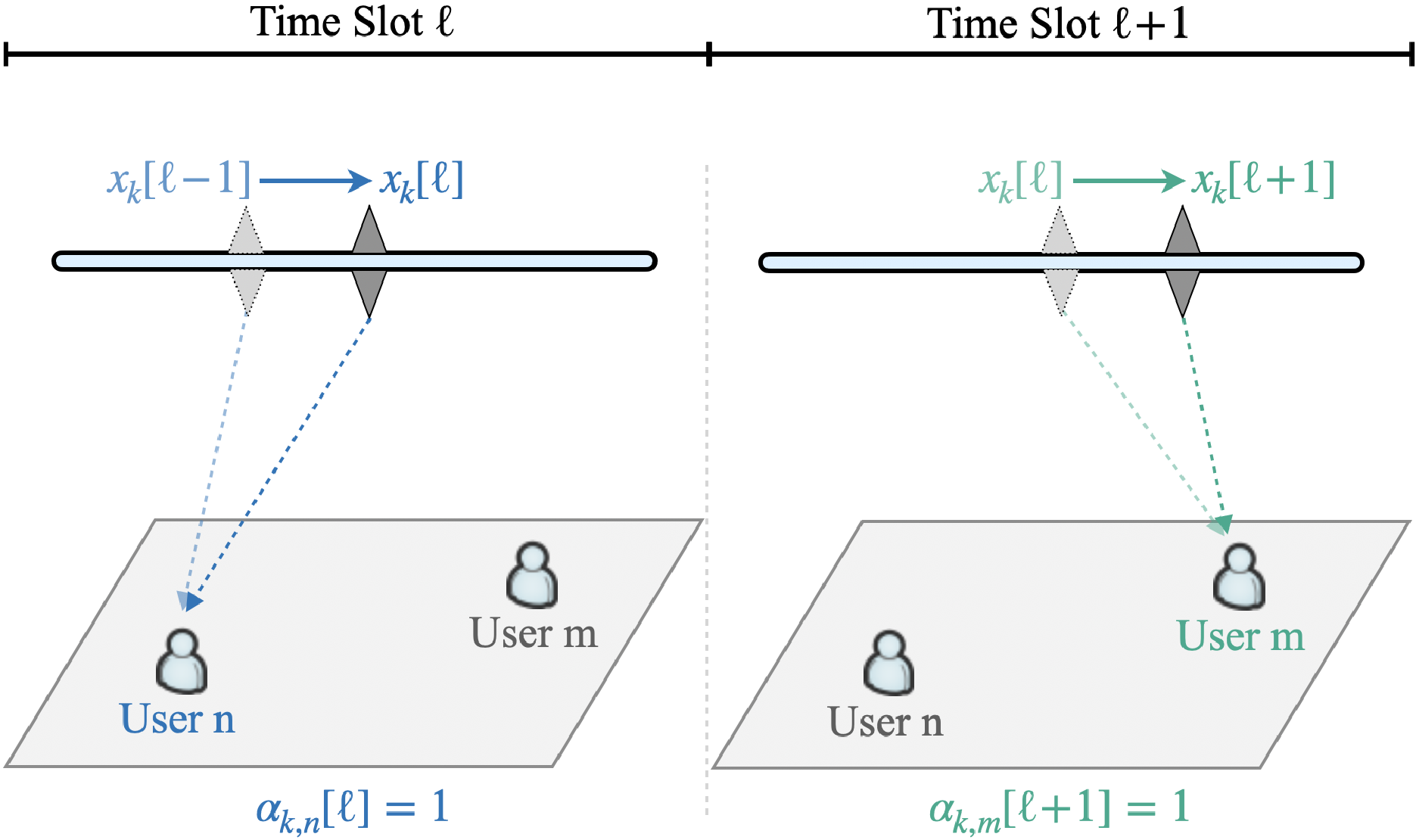}
\caption{Illustration of trajectory based transmission and user scheduling. During time slot $\ell$, the pinching antenna moves toward user $m$ while serving user $n$. In time slot $\ell+1$, it continues moving while serving user $m$.}\vspace{-4mm}
\label{system}
\end{figure}

TDMA is adopted in the considered system, with one user scheduled for transmission in each time slot. A service period of duration $T$ is uniformly divided into $L$ time slots, indexed by $\mathcal{L}=\{1,2,\cdots,L\}$, each with duration $\Delta t=T/L$. During the service period, the pinching antennas can move continuously along their corresponding waveguides, with downlink transmission maintained by the active antenna. In the considered trajectory model, each antenna's position in time slot $\ell$ is represented by its position at the end of that slot, as shown in \fref{system}. Accordingly, the position of pinching antenna $k$ in time slot $\ell$ is  expressed as
\begin{equation}
\boldsymbol{\psi}_k^\mathrm{pin}[\ell]=(x_k[\ell],y_k,d),
\end{equation}
where $x_k[\ell]$ is the $x$-coordinate of pinching antenna $k$ in time slot $\ell$ and satisfies $-D_x/2\leq x_k[\ell]\leq D_x/2$, $\forall k\in\mathcal{K}$, $\ell\in\mathcal{L}$.

To ensure periodic operation, antenna trajectories are planned such that each pinching antenna returns to its initial position by the end of the service period while satisfying the movement speed constraint, as follows:
\begin{equation}
x_k[L]=x_k[0],\,\forall k\in\mathcal{K},
\end{equation}
where $x_k[0]$ is the initial $x$-coordinate of pinching antenna $k$. Moreover, the displacement of each pinching antenna between consecutive time slots is limited by the maximum movement speed $v_\mathrm{max}$, leading to the following constraint:
\begin{equation}
\left|x_k[\ell]-x_k[\ell\!-\!1]\right|\leq v_\mathrm{max}\Delta t,\,\forall k\in\mathcal{K},\,\ell\in\mathcal{L}.
\end{equation}
Note that the effects of antenna acceleration and Doppler shifts during movement are neglected\footnote{In the considered model, the pinching antennas move at low speeds along preplanned trajectories. The corresponding phase variations can be predicted from the antenna motion and propagation geometry, facilitating phase tracking and compensation through synchronization and channel estimation \cite{kaidi2026roaming}.}.

In \cite{kaidi2026roaming}, antenna placement and antenna roaming were investigated as two operating modes for pinching-antenna systems. The relationship between these operating modes and the proposed scheme is discussed in the following remark.
\begin{remark}
The proposed trajectory based scheme can be regarded as a combination of antenna placement and antenna roaming in \cite{kaidi2026roaming}. By permitting transmission during both antenna movement and stationary operation, the proposed scheme provides greater operational flexibility, with the potential to further improve communication performance.
\end{remark}
%%%%%%%%%%%%%%%%%%%%%%%%%%%%%%%%%%%%%%%%%%%%%%%%%
\subsection{Signal Model}
An LoS dominated channel is considered between each pinching antenna and user, while the propagation loss within the dielectric waveguides is assumed to be negligible \cite{xu2026loss}. By incorporating the phase shifts introduced by both in-waveguide and free-space propagation, the channel coefficient between pinching antenna $k$ and user $n$ during time slot $\ell$ is given by
\begin{equation}
h_{k,n}[\ell]=\frac{\eta e^{-j\left(\frac{2\pi}{\lambda}\left\|\boldsymbol{\psi}_n-\boldsymbol{\psi}_k^\mathrm{pin}[\ell]\right\|+\frac{2\pi}{\lambda_\mathrm{g}}\left\|\boldsymbol{\psi}_k^\mathrm{feed}-\boldsymbol{\psi}_k^\mathrm{pin}[\ell]\right\|\right)}}{\left\|\boldsymbol{\psi}_n-\boldsymbol{\psi}_k^\mathrm{pin}[\ell]\right\|},
\end{equation}
where $\eta=\lambda/(4\pi)$, $\lambda$ is the carrier wavelength in free space, $\lambda_\mathrm{g}=\lambda/n_\mathrm{eff}$ is the guided wavelength in the waveguide, $n_\mathrm{eff}$ is the effective refractive index of the dielectric waveguide, and $\|\cdot\|$ denotes the Euclidean norm.

To characterize the antenna-user service association in each time slot, a binary user scheduling indicator $\alpha_{k,n}[\ell]\in\{0,1\}$ is introduced, where $\alpha_{k,n}[\ell]=1$ indicates that user $n$ is served by pinching antenna $k$ during time slot $\ell$, and $\alpha_{k,n}[\ell]=0$ otherwise, as shown in \fref{system}. Under the adopted TDMA protocol, exactly one antenna-user pair is selected for transmission in each time slot. Accordingly, the scheduling indicators satisfy
\begin{equation}
\sum_{k=1}^{K}\sum_{n=1}^{N}\alpha_{k,n}[\ell]=1,\,\forall\ell\in\mathcal{L}.
\end{equation}
That is, all pinching antennas other than the selected one remain inactive. The advantage of deploying multiple waveguides is discussed in the following remark.
\begin{remark}\label{multiplepin}
Although only one pinching antenna is active for transmission in each time slot, antennas on the other waveguides can simultaneously move toward favorable positions for subsequent user service. This parallel repositioning allows the system to benefit from multiple waveguides by selecting suitably positioned antennas across time slots.
\end{remark}

During each time slot, the transmit power $P_\mathrm{t}$ is allocated to the active pinching antenna. When user $n$ is scheduled in time slot $\ell$, its received signal can be expressed as follows:
\begin{equation}
y_n[\ell]=\sum_{k=1}^{K}\alpha_{k,n}[\ell]h_{k,n}[\ell]\sqrt{P_\mathrm{t}}s_n[\ell]+z_n[\ell],
\end{equation}
where $s_n[\ell]$ is the desired signal of user $n$, and $z_n[\ell]$ is the additive noise. The average achievable rate\footnote{Since the pinching antenna may move within a slot, the slot averaged achievable rate is given by the integral of the instantaneous rate over that slot, normalized by $\Delta t$. Accordingly, \eqref{slotrate} provides an approximation to this average, with the approximation error tending to zero as $\Delta t \to 0$ \cite{wu2018uav, wu2018uav2}.} of user $n$ in time slot $\ell$, measured in bit/s/Hz, is approximated as
\begin{equation}\label{slotrate}
r_n[\ell]=\sum_{k=1}^{K}\alpha_{k,n}[\ell]\log_2\!\left(1+\frac{P_\mathrm{t}\!\left|h_{k,n}[\ell]\right|^2}{\sigma^2}\right),
\end{equation}
where $\sigma^2$ is the noise power. Therefore, the average achievable data rate of user $n$ over the service period is
\begin{equation}
R_n=\frac{1}{L}\!\sum_{\ell=1}^{L}r_n[\ell]=\frac{1}{L}\sum_{\ell=1}^{L}\!\sum_{k=1}^{K}\alpha_{k,n}[\ell]\log_2\!\left(\!1\!+\!\frac{P_\mathrm{t}\!\left|h_{k,n}[\ell]\right|^2}{\sigma^2}\!\right).
\end{equation}
In the considered system, the number of time slots is greater than the number of users, i.e., $L>N$, and hence each user may be allocated multiple time slots within a service period.
%%%%%%%%%%%%%%%%%%%%%%%%%%%%%%%%%%%%%%%%%%%%%%%%%
%%%%%%%%%%%%%%%%%%%%%%%%%%%%%%%%%%%%%%%%%%%%%%%%%
\section{Performance Analysis}
To characterize the performance of the proposed scheme, a simplified single-waveguide scenario is considered, where one pinching antenna, indexed by $k$, serves $N$ users. The waveguide is located at $y_k=0$, and the antenna is initially positioned at $x_k[0]=0$. To facilitate analysis, a continuous-time representation of antenna trajectories and user scheduling is adopted, corresponding to the limit of the time slotted model as $\Delta t\to 0$. The users are independently and uniformly distributed over the service region, with horizontal coordinates ordered as $x_{(1)}\leq\cdots\leq x_{(N)}$. The pinching antenna serves users along the following trajectory:
\begin{equation}
0 \rightarrow x_{(1)}\rightarrow \cdots \rightarrow x_{(N)} \rightarrow 0,
\end{equation}
moving at speed $v_\mathrm{max}$ between stationary service periods. During antenna movement, the user with the smallest horizontal distance to the antenna is served, with service switching at the midpoint between adjacent user coordinates. The service boundaries are defined as follows:
\begin{equation}
b_i=\begin{cases}
\min\{0,x_{(1)}\}, & \text{if }i=0,\\
\dfrac{x_{(i)}+x_{(i+1)}}{2}, & \text{if }i=1,\cdots,N-1,\\
\max\{0,x_{(N)}\}, & \text{if }i=N.
\end{cases}
\end{equation}
Hence, the user at $x_{(i)}$ is served while the antenna moves within $[b_{i-1},b_i]$. The time available for stationary transmission, obtained by subtracting the total antenna movement time from the service period duration, is allocated equally among all users. During stationary transmission to user $n$, the antenna is positioned at the point on the waveguide closest to that user, i.e., $x_k=x_n$. The achievable data rate of user $n$ at antenna position $x_k$, averaged over the transverse user coordinate $y_n$ for a given horizontal coordinate $x_n$, can be expressed as
\begin{equation}\label{barr}
\bar{r}_n(x_k)=\frac{1}{D_y}\int_{-\frac{D_y}{2}}^\frac{D_y}{2}\!\log_2\!\left(1+\frac{\rho}{d^2+y_n^2+(x_k-x_n)^2}\right)\mathrm{d}y_n,
\end{equation}
where $\rho=P_\mathrm{t}\eta^2/\sigma^2$. The expected average achievable rate of each user is characterized in the following proposition.
\begin{proposition}\label{expectrate}
Under $v_\mathrm{max}T\geq2D_x$, the expected average achievable rate of user $n$ over the service period is given by
\begin{align}\label{expectr}
\bar{R}_n^\mathrm{traj}&=\frac{1}{N}\left[1-\frac{2D_x(N-1+2^{-N})}{(N+1)v_\mathrm{max}T}\right]\bar{r}_n(x_n)\\
&\quad+\frac{2(N-1+2^{-N})}{Nv_\mathrm{max}T}\!\!\int_{-\frac{D_x}{2}}^\frac{D_x}{2}\!\!\left(1-\frac{2|z|}{D_x}\right)^N\!\bar{r}_n(x_n\!+\!z)\mathrm{d}z.\nonumber
\end{align}
\end{proposition}
\begin{IEEEproof}
Refer to Appendix~A.
\end{IEEEproof}

The following remark provides further insights into \eqref{expectr}.
\begin{remark}
The expected average rate in \eqref{expectr} consists of two components. The first term represents the contribution from stationary transmission at the position closest to user $n$, i.e., $x_k=x_n$. The second term represents the contribution from transmission to user $n$ while the antenna moves along the waveguide, with $z=x_k-x_n$ denoting the signed displacement of the antenna from the user along the $x$-axis.
\end{remark}

To evaluate the performance of the trajectory based scheme, the conventional antenna placement scheme with instantaneous positioning is considered as a benchmark. In this scheme, the antenna is placed at the position closest to the scheduled user, i.e., $x_k=x_n$, with negligible repositioning time, thereby maximizing the achievable rate during transmission \cite{xie2025pin, ding2025analytical}. With equal time allocation, the expected average achievable rate of user $n$ is given by
\begin{equation}\label{expectrinst}
\bar{R}_n^\mathrm{inst}=\frac{1}{N}\bar{r}_n(x_n)=\frac{1}{ND_y}\!\int_{-\frac{D_y}{2}}^\frac{D_y}{2}\!\log_2\!\left(1+\frac{\rho}{d^2+y_n^2}\right)\mathrm{d}y_n.
\end{equation}
The following proposition characterizes the rate loss relative to instantaneous positioning.
\begin{proposition}\label{compinst}
Under $v_\mathrm{max}T\geq2D_x$, the expected rate loss relative to instantaneous positioning, defined as $\Delta_n^\mathrm{inst}=\bar{R}_n^\mathrm{inst}-\bar{R}_n^\mathrm{traj}$, is given by
\begin{equation}\label{gapinst}
\Delta_n^\mathrm{inst}\!=\!\frac{2(N\!-\!1\!+\!2^{-N})}{Nv_\mathrm{max}T}\!\!\!\int_{-\frac{D_x}{2}}^\frac{D_x}{2}\!\!\!\left(\!\!1\!-\!\frac{2|z|}{D_x}\!\right)^N\!\!\![\bar{r}_n(x_n)-\bar{r}_n(x_n\!+z)]\mathrm{d}z.
\end{equation}
For $\rho>0$ and finite $v_\mathrm{max}T$, it satisfies
\begin{equation}\label{gapinstupper}
0<\Delta_n^\mathrm{inst}\leq\frac{\mu}{N}\!\left[\bar{r}_n(x_n)-\bar{r}_n\!\!\left(x_n\!+\!\frac{D_x}{\sqrt{2(N+2)(N+3)}}\right)\right],
\end{equation}
where $\mu=\frac{2D_x(N-1+2^{-N})}{(N+1)v_\mathrm{max}T}$ is the expected fraction of the service period spent moving.
\end{proposition}
\begin{IEEEproof}
Refer to Appendix~B.
\end{IEEEproof}

Further insights into Proposition~\ref{compinst} are provided below.
\begin{remark}
The expected rate loss relative to instantaneous positioning depends on both the expected movement-time fraction $\mu$ and the channel quality degradation during movement. Since transmission continues during movement, the loss in \eqref{gapinst} arises from the nonzero antenna-user distance $|z|$ along the $x$-axis. Furthermore, the coefficient multiplying the integral in \eqref{gapinst} equals $\mu(N+1)/(ND_x)$, indicating that $\mu$ scales the integrated rate reduction.
\end{remark}
\begin{remark}
With other parameters fixed, the expected rate loss in \eqref{gapinst} is inversely proportional to $v_\mathrm{max}T$. Therefore, increasing $v_\mathrm{max}$ or $T$ improves the expected average rate of the proposed trajectory based scheme, which approaches the instantaneous-positioning benchmark as $v_\mathrm{max}T\to\infty$. In contrast, increasing $D_x$ expands the integration interval and increases the weighting factor $(1-2|z|/D_x)^N$ in \eqref{gapinst}, thereby increasing the rate loss and reducing the expected average rate, provided that the feasibility condition remains satisfied.
\end{remark}

A fixed-antenna scheme is also considered as a benchmark, where the antenna is located at $(0,0,d)$ and serves all users with equal time allocation. According to \cite{ding2024pin}, the expected average achievable rate of user $n$ is given by
\begin{equation}\label{expectrfix}
\bar{R}_n^\mathrm{fix}=\frac{1}{ND_xD_y}\!\int_{-\frac{D_x}{2}}^{\frac{D_x}{2}}\!\int_{-\frac{D_y}{2}}^\frac{D_y}{2}\!\log_2\!\left(1\!+\!\frac{\rho}{d^2\!+\!y_n^2\!+\!x_n^2}\right)\mathrm{d}y_n\mathrm{d}x_n.
\end{equation}
The following proposition characterizes the rate improvement over the fixed-antenna scheme.
\begin{proposition}\label{compfix}
Under $v_\mathrm{max}T\geq2D_x$, the expected rate improvement over the fixed-antenna scheme, defined as $\Delta_n^\mathrm{fix}=\bar{R}_n^\mathrm{traj}-\bar{R}_n^\mathrm{fix}$, is given by
\begin{align}\label{gapfix}
\Delta_n^\mathrm{fix}\!&=\!\frac{1}{ND_x}\int_{-\frac{D_x}{2}}^\frac{D_x}{2}[\bar{r}_n(x_n)\!-\!\bar{r}_n(x_n\!+\!z)]\mathrm{d}z\\
&\!\!\!\!\!\!-\!\frac{2(N\!-\!1\!+\!2^{-N})}{Nv_\mathrm{max}T}\!\!\!\int_{-\frac{D_x}{2}}^\frac{D_x}{2}\!\!\left(\!1\!-\!\frac{2|z|}{D_x}\!\right)^N\!\![\bar{r}_n(x_n)\!-\!\bar{r}_n(x_n\!+\!z)]\mathrm{d}z.\nonumber
\end{align}
For $\rho>0$, it satisfies the strictly positive lower bound
\begin{align}\label{gapfixlower}
\Delta_n^\mathrm{fix}&\geq\frac{\rho D_x^2}{12N\ln(2)\!\left(d^2\!+\!\frac{D_y^2}{12}\!+\!\frac{D_x^2}{4}\right)\left(d^2\!+\!\frac{D_y^2}{12}\!+\!\frac{D_x^2}{4}\!+\!\rho\right)}
\nonumber\\
&\quad\times\left[1-\frac{6\mu}{(N+2)(N+3)}\right]>0,
\end{align}
where $\mu$ is defined in Proposition~\ref{compinst}.
\end{proposition}
\begin{IEEEproof}
Refer to Appendix~C.
\end{IEEEproof}

Further insights into Proposition~\ref{compfix} are provided below.
\begin{remark}
By adapting the antenna position to user locations, the proposed trajectory based scheme achieves a strictly higher expected average rate than the fixed-antenna scheme. With other parameters fixed, increasing $v_\mathrm{max}$ or $T$ further increases the expected rate improvement. This improvement approaches $\bar{R}_n^\mathrm{inst}-\bar{R}_n^\mathrm{fix}$ as $v_\mathrm{max}T\to\infty$.
\end{remark}
%%%%%%%%%%%%%%%%%%%%%%%%%%%%%%%%%%%%%%%%%%%%%%%%%
%%%%%%%%%%%%%%%%%%%%%%%%%%%%%%%%%%%%%%%%%%%%%%%%%
\section{Problem Formulation}
Due to differences in user locations and the limited antenna movement speed, users may experience different channel conditions and service opportunities over the service period. To improve user fairness, the minimum average achievable rate among all users is adopted as the performance metric. Accordingly, the antenna trajectories and user scheduling are jointly optimized as follows:
\begin{subequations}\label{problem}
\begin{empheq}{align}
\max_{\boldsymbol{x},\boldsymbol{\alpha}}\quad & \min\{R_n|\forall n\in\mathcal{N}\}\\
\textrm{s.t.}\quad 
& x_k[L]=x_k[0],\;\forall k\in\mathcal{K},\\
& |x_k[\ell]\!-\!x_k[\ell\!-\!1]|\le v_\mathrm{max}\Delta t,\;\forall k\in\mathcal{K},\;\forall \ell\in\mathcal{L},\\
& -D_x/2\le x_k[\ell]\le D_x/2,\,\forall k\in\mathcal{K},\;\forall\ell\in\mathcal{L},\\
& \alpha_{k,n}[\ell]\in\{0,1\},\,\forall k\in\mathcal{K},\;\forall n\in\mathcal{N},\;\forall \ell\in\mathcal{L},\\
& \sum\nolimits_{k=1}^{K}\sum\nolimits_{n=1}^{N}\!\alpha_{k,n}[\ell]= 1,\;\forall\ell\in\mathcal L,
\end{empheq}
\end{subequations}\vspace{-2mm}\\
where $\boldsymbol{x}=\{x_k[\ell]\}_{k\in\mathcal{K},\ell\in\mathcal{L}}$ and $\boldsymbol{\alpha}=\{\alpha_{k,n}[\ell]\}_{k\in\mathcal{K},n\in\mathcal{N},\ell\in\mathcal{L}}$ are the collections of all antenna positions and user scheduling indicators, respectively. Constraint (\ref{problem}b) ensures that each pinching antenna returns to its prescribed initial position at the end of the service period. Constraint (\ref{problem}c) limits the antenna displacement between consecutive sampling instants according to the maximum movement speed $v_\mathrm{max}$, while constraint (\ref{problem}d) confines each pinching antenna to be within the waveguide length. Constraint (\ref{problem}e) specifies the binary user scheduling indicators, and constraint (\ref{problem}f) guarantees that exactly one antenna is selected to serve one user in each time slot.

Problem \eqref{problem} is a mixed-integer non-convex optimization problem due to the binary user scheduling indicators and the non-concave dependence of the achievable rates on the antenna positions. Moreover, the antenna trajectories and user scheduling indicators are coupled through the average achievable rates. To facilitate the solution, an AO framework is adopted, in which problem \eqref{problem} is decomposed into two subproblems, namely antenna trajectory design and user scheduling. Given the scheduling solution $\boldsymbol{\alpha}^{(t)}$ at iteration $t$, the antenna trajectories are optimized by solving
\begin{subequations}\label{tdproblem}
\begin{empheq}{align}
\max_{\boldsymbol{x}}\quad & \min\{R_n|\forall n\in\mathcal{N}\}\\
\mathrm{s.t.}\quad & \text{(\ref{problem}b)}, \text{(\ref{problem}c)}, \text{and (\ref{problem}d)}.\nonumber
\end{empheq}
\end{subequations}\vspace{-2mm}\\
With the updated antenna trajectories $\boldsymbol{x}^{(t+1)}$, the user scheduling indicators are optimized by solving
\begin{subequations}\label{usproblem}
\begin{empheq}{align}
\max_{\boldsymbol{\alpha}}\quad & \min\{R_n|\forall n\in\mathcal{N}\}\\
\mathrm{s.t.}\quad & \text{(\ref{problem}e)}, \text{and (\ref{problem}f)}.\nonumber
\end{empheq}
\end{subequations}\vspace{-2mm}\\
The resulting scheduling solution $\boldsymbol{\alpha}^{(t+1)}$ is then used for trajectory design in the next iteration. These two steps are repeated until convergence.
%%%%%%%%%%%%%%%%%%%%%%%%%%%%%%%%%%%%%%%%%%%%%%%%%
%%%%%%%%%%%%%%%%%%%%%%%%%%%%%%%%%%%%%%%%%%%%%%%%%
\section{SCA based Trajectory Design}
In this section, the antenna trajectories are optimized for a given user scheduling solution. Accordingly, the active pinching antenna and the scheduled user in each time slot are held fixed during the trajectory update, with transmit power allocated to the active antenna. By introducing an auxiliary variable $\delta$, problem \eqref{tdproblem} can be equivalently rewritten as
\begin{subequations}\label{tdproblem1}
\begin{empheq}{align}
\max_{\boldsymbol{x}, \delta}\quad & \delta\\
\mathrm{s.t.}\quad & R_n \ge \delta,\;\forall n \in\mathcal{N},\\
& \text{(\ref{problem}b)}, \text{(\ref{problem}c)}, \text{and (\ref{problem}d)}.\nonumber
\end{empheq}
\end{subequations}\vspace{-2mm}\\
Problem \eqref{tdproblem1} remains non-convex because the average achievable rates are non-concave functions of the antenna positions. To address this issue, the SCA method is applied.

Based on the considered channel model, the channel power gain between antenna $k$ and user $n$ in time slot $\ell$ is given by
\begin{equation}
\left|h_{k,n}[\ell]\right|^2=\frac{\eta^2}{(x_k[\ell]\!-\!x_n)^2\!+\!(y_k\!-\!y_n)^2\!+\!d^2}.
\end{equation}
When user $n$ is served by pinching antenna $k$, its achievable rate in time slot $\ell$ can be expressed as follows:
\begin{equation}
r_{k,n}[\ell]=\log_2\!\left(1\!+\!\frac{\rho}{(x_k[\ell]\!-\!x_n)^2\!+\!(y_k\!-\!y_n)^2\!+\!d^2}\right).
\end{equation}
Therefore, the average achievable rate of user $n$ over the service period is given by
\begin{equation}
R_n=\frac{1}{L}\sum_{\ell=1}^{L}\sum_{k=1}^{K}\alpha_{k,n}[\ell]r_{k,n}[\ell].
\end{equation}

For each time slot $\ell$, antenna $k$, and user $n$, $r_{k,n}[\ell]$ is convex with respect to the squared horizontal distance $(x_k[\ell]-x_n)^2$. Hence, its first-order Taylor approximation with respect to this squared distance provides a global lower bound. At the $i$-th SCA iteration, define
\begin{equation}
q_{k,n}^{(i)}[\ell]\triangleq(x_k^{(i)}[\ell]-x_n)^2+(y_k-y_n)^2+d^2,
\end{equation}
where $x_k^{(i)}[\ell]$ is the $x$-coordinate of pinching antenna $k$ in time slot $\ell$ at the $i$-th SCA iteration. The negative of the derivative with respect to the squared horizontal distance, evaluated at the current trajectory, is given by
\begin{equation}
A_{k,n}^{(i)}[\ell]=\frac{\rho}{\ln(2)q_{k,n}^{(i)}[\ell](q_{k,n}^{(i)}[\ell]+\rho)}.
\end{equation}
Accordingly, the achievable rate is lower bounded as $r_{k,n}[\ell]\geq\underline{r}_{k,n}^{(i)}[\ell](x_k[\ell])$, where
\begin{align}
\underline{r}_{k,n}^{(i)}[\ell](x_k[\ell])&\triangleq\log_2\!\left(1+\frac{\rho}{q_{k,n}^{(i)}[\ell]}\right)\\
&\quad-A_{k,n}^{(i)}[\ell]\!\left[(x_k[\ell]\!-\!x_n)^2\!-\!(x_k^{(i)}[\ell]\!-\!x_n)^2\right].\nonumber
\end{align}
Since $A_{k,n}^{(i)}[\ell]\geq0$, the constructed lower bound is concave in $x_k[\ell]$ and is tight at $x_k[\ell]=x_k^{(i)}[\ell]$. Based on the concave lower bound of the instantaneous rate, the average rate lower bound of user $n$ at the $i$-th SCA iteration is defined as
\begin{equation}
\underline{R}_n^{(i)}(\boldsymbol{x})\triangleq\frac{1}{L}\sum_{\ell=1}^{L}\sum_{k=1}^{K}\alpha_{k,n}[\ell]\underline{r}_{k,n}^{(i)}[\ell](x_k[\ell]).
\end{equation}
Since the user scheduling indicators are fixed and nonnegative, $\underline{R}_n^{(i)}(\boldsymbol{x})$ is concave in $\boldsymbol{x}$ and satisfies
\begin{equation}
\underline{R}_n^{(i)}(\boldsymbol{x})\leq R_n(\boldsymbol{x}),
\end{equation}
and
\begin{equation}
\underline{R}_n^{(i)}(\boldsymbol{x}^{(i)})=R_n(\boldsymbol{x}^{(i)}).
\end{equation}
Therefore, the constructed approximation provides a global concave lower bound of the average achievable rate and is tight at the current trajectory. At the $i$-th SCA iteration, the antenna trajectories are updated by solving the following problem:
\begin{subequations}\label{tdproblem2}
\begin{empheq}{align}
\max_{\boldsymbol{x},\delta}\quad
& \delta\\
\mathrm{s.t.}\quad
& \underline{R}_n^{(i)}(\boldsymbol{x})\geq\delta,\;\forall n\in\mathcal N,\\
& \text{(\ref{problem}b)}, \text{(\ref{problem}c)}, \text{and (\ref{problem}d)}.\nonumber
\nonumber
\end{empheq}
\end{subequations}\vspace{-2mm}\\
Since $\underline{R}_n^{(i)}(\boldsymbol{x})$ is a concave quadratic function, constraint (\ref{tdproblem2}b) defines a convex feasible set in $(\boldsymbol{x},\delta)$. Moreover, the antenna position, movement, and periodicity constraints are convex. Hence, problem \eqref{tdproblem2} is a convex quadratically constrained quadratic program (QCQP), which can be efficiently solved using standard convex optimization solvers.

\begin{algorithm}[t]
\caption{SCA based Trajectory Design Algorithm}
\label{tdalg}
\begin{algorithmic}[1]
\STATE Set $i=0$ with a given feasible trajectory $\boldsymbol{x}^{(0)}$.
\STATE Calculate $\delta^{(0)}=\min_{n\in\mathcal{N}}R_n(\boldsymbol{x}^{(0)})$.
\REPEAT
\STATE Calculate $q_{k,n}^{(i)}[\ell]$ and $A_{k,n}^{(i)}[\ell]$ based on $\boldsymbol{x}^{(i)}$.
\STATE Construct $\underline{r}_{k,n}^{(i)}[\ell](x_k[\ell])$ and $\underline{R}_n^{(i)}(\boldsymbol{x})$.
\STATE Solve problem \eqref{tdproblem2} to obtain $(\boldsymbol{x}^{\star(i)},\delta^{\star(i)})$.
\STATE Update $\boldsymbol{x}^{(i+1)}=\boldsymbol{x}^{\star(i)}$.
\STATE Calculate $\delta^{(i+1)}=\min_{n\in\mathcal{N}}R_n(\boldsymbol{x}^{(i+1)})$.
\STATE Set $i\leftarrow i+1$.
\UNTIL{$\delta^{(i)}-\delta^{(i-1)}\leq\epsilon$ or $i\geq I_\mathrm{max}$}
\RETURN $\boldsymbol{x}^{(i)}$.
\end{algorithmic}
\end{algorithm}

Based on the preceding convex approximation, the SCA based trajectory design procedure is summarized in Algorithm~\ref{tdalg}. The convergence tolerance and maximum number of iterations are denoted by $\epsilon>0$ and $I_\mathrm{max}$, respectively. For the first AO iteration, each user is associated with its nearest waveguide to construct the initial trajectories. With associated users on one side of $x_k[0]$, the antenna moves toward the farthest user and reverses direction at $T/2$. Its displacement from $x_k[0]$ at reversal equals the smaller of that user's horizontal distance from $x_k[0]$ and $v_\mathrm{max}T/2$. With users on both sides, the antenna reaches $x_k[0]-A_k$ and $x_k[0]+A_k$ at $T/4$ and $3T/4$, respectively, where $A_k=\min\{D_x/2-|x_k[0]|,v_\mathrm{max}T/4\}$. The antennas move at a constant speed along each segment and return to $x_k[0]$ at $T$, with the trajectories discretized at intervals of $\Delta t$. Antennas without associated users remain stationary. For subsequent executions, the trajectories from the preceding AO iteration are employed for initialization. At each SCA iteration, the rate lower bounds are constructed at $\boldsymbol{x}^{(i)}$, and problem \eqref{tdproblem2} is solved to obtain $(\boldsymbol{x}^{\star(i)},\delta^{\star(i)})$. The updated trajectory $\boldsymbol{x}^{(i+1)}=\boldsymbol{x}^{\star(i)}$ serves as the reference point for the next iteration. While $\delta^{\star(i)}$ is the optimal objective value of the convex approximation, $\delta^{(i+1)}$ is the actual minimum average rate at the updated trajectory. The improvement in this actual rate is used to determine convergence, subject to the maximum iteration limit.

The convergence of the objective value in Algorithm~\ref{tdalg} is analyzed below. Since the rate lower bounds are tight at the current trajectory, it follows that $\underline{R}_n^{(i)}(\boldsymbol{x}^{(i)})=R_n(\boldsymbol{x}^{(i)})$ for all $n\in\mathcal{N}$. Thus, $(\boldsymbol{x}^{(i)},\delta^{(i)})$ is feasible for problem \eqref{tdproblem2}, and its optimal objective value satisfies $\delta^{\star(i)}\geq\delta^{(i)}$. Moreover, since the approximate rates are global lower bounds of the actual rates and $\boldsymbol{x}^{(i+1)}=\boldsymbol{x}^{\star(i)}$, the updated minimum average rate satisfies the following condition:
\begin{align}
\delta^{(i+1)}&=\min_{n\in\mathcal{N}}R_n(\boldsymbol{x}^{(i+1)})\nonumber\\
&\geq\min_{n\in\mathcal{N}}\underline{R}_n^{(i)}(\boldsymbol{x}^{\star(i)})\geq\delta^{\star(i)}\geq\delta^{(i)}.
\end{align}
Therefore, the actual minimum average rate is non-decreasing over the SCA iterations. Since the achievable rates are upper bounded, the objective value sequence $\{\delta^{(i)}\}$ converges.
%%%%%%%%%%%%%%%%%%%%%%%%%%%%%%%%%%%%%%%%%%%%%%%%%
%%%%%%%%%%%%%%%%%%%%%%%%%%%%%%%%%%%%%%%%%%%%%%%%%
\section{Game Theory based User Scheduling}
In this section, the user scheduling problem in \eqref{usproblem} is investigated for given antenna trajectories. Since the scheduling decisions across time slots are coupled through the average achievable rates, a scheduling game $(\mathcal{L},\mathcal{S},\boldsymbol{u})$ is formulated, where the time slots in $\mathcal{L}$ are treated as the players, $\mathcal{S}$ is the current scheduling configuration, and $\boldsymbol{u}(\mathcal{S})$ is the common utility. The scheduling state of player $\ell$ is denoted by $s_\ell=(k,n)$, indicating that pinching antenna $k$ serves user $n$ in time slot $\ell$. Accordingly, the current scheduling configuration is defined as follows:
\begin{equation}
\mathcal{S}\triangleq\left\{\langle\ell,s_\ell\rangle\,|\,\forall \ell\in\mathcal{L}\right\}.
\end{equation}
Each configuration determines the scheduling indicators through $\alpha_{k,n}[\ell]=1$ if $s_\ell=(k,n)$ and $\alpha_{k,n}[\ell]=0$ otherwise. Since each player selects exactly one antenna-user pair, the resulting scheduling indicators satisfy both the binary constraint (\ref{problem}e) and the single-pair scheduling constraint (\ref{problem}f).

The formulated game exhibits externalities because changing the scheduling state of one player affects the average user rates and hence the common utility of all players. In this case, directly using the minimum average achievable rate as the utility may ignore beneficial scheduling updates if only strict utility improvements are accepted. To distinguish configurations with the same minimum rate while retaining it as the primary objective, the common utility is defined as the ordered user-rate vector, as follows:
\begin{equation}\label{utility}
\boldsymbol{u}(\mathcal{S})\triangleq\left[R_{(1)}(\mathcal{S}),R_{(2)}(\mathcal{S}),\cdots,R_{(N)}(\mathcal{S})\right],
\end{equation}
where $R_{(1)}(\mathcal{S})\leq R_{(2)}(\mathcal{S})\leq\cdots\leq R_{(N)}(\mathcal{S})$ are the average achievable rates under scheduling configuration $\mathcal{S}$, arranged in ascending order. For two configurations $\mathcal{S}$ and $\mathcal{S}'$, their utility vectors are compared sequentially from the first component. Specifically, $\mathcal{S}$ is preferred to $\mathcal{S}'$ if and only if there exists an index $n\in\mathcal{N}$ such that the following conditions hold:
\begin{equation}\label{preference}
\mathcal{S}\succ\mathcal{S}'\Longleftrightarrow
\begin{cases}
R_{(i)}(\mathcal{S})=R_{(i)}(\mathcal{S}'), & \forall i<n,\\
R_{(n)}(\mathcal{S})>R_{(n)}(\mathcal{S}'). &
\end{cases}
\end{equation}
That is, a configuration with a higher minimum average rate is preferred. When the minimum rates are equal, the remaining ordered rates are compared sequentially. Since all players share the same utility vector, their preferences are aligned.

To explore the scheduling configurations efficiently, two update stages are adopted. The first stage allows each player to change its antenna-user pair unilaterally. The second stage exchanges the antenna-user pairs assigned to two players\footnote{Such coordinated exchanges may improve the utility even when unilateral updates cannot. For example, consider a single-antenna system serving two users with equal average rates. Suppose two of the $L$ time slots are assigned to different users, and each user can achieve a higher rate in the other slot. Reassigning either slot alone reduces the average rate of the user losing that slot and hence the minimum rate. In contrast, exchanging the assignments increases both users' average rates.}. The two stages are described as follows:
\begin{enumerate}[leftmargin=*]
\item \textit{Pair Construction Stage:} In this stage, player $\ell$ replaces its current state $s_\ell$ with a candidate state $s_\ell'$. The resulting scheduling configuration  is given by
\begin{equation}\label{construction}
\mathcal{S}'=\mathcal{S}\setminus\left\{\langle\ell,s_\ell\rangle\right\}\cup\left\{\langle\ell,s_\ell'\rangle\right\}.
\end{equation}
The current configuration is updated as $\mathcal{S}\leftarrow\mathcal{S}'$ if $\mathcal{S}'\succ\mathcal{S}$; otherwise, it remains unchanged.
\item \textit{Pair Exchange Stage:} In this stage, player $\ell$ proposes to exchange its antenna-user pair with player $i\in\mathcal{L}$, where $i\neq\ell$. The candidate scheduling configuration is given by
\begin{equation}\label{exchange}
\mathcal{S}'=\mathcal{S}\setminus\left\{\langle\ell,s_\ell\rangle,\langle i,s_i\rangle\right\}\cup\left\{\langle\ell,s_i\rangle,\langle i,s_\ell\rangle\right\}.
\end{equation}
Since all players share the same utility, player $i$ accepts the proposal if $\mathcal{S}'\succ\mathcal{S}$, and the configuration is updated as $\mathcal{S}\leftarrow\mathcal{S}'$. Otherwise, the proposal is rejected and the configuration remains unchanged.
\end{enumerate}

Based on the preceding two-stage updates, the stability of a scheduling configuration is defined as follows:
\begin{definition}
A feasible scheduling configuration $\mathcal{S}$ is stable with respect to the proposed two-stage updates if
\begin{equation}
\mathcal{S}'\not\succ\mathcal{S}
\end{equation}
for every feasible candidate configuration $\mathcal{S}'$ obtained from $\mathcal{S}$ through either pair construction or pair exchange.
\end{definition}
At a stable configuration, no unilateral pair construction or bilateral pair exchange can improve the utility, corresponding to Nash stability and exchange stability, respectively \cite{han2012game}.

\begin{algorithm}[t]
\caption{Game based User Scheduling Algorithm}
\label{usalg}
\begin{algorithmic}[1]
\STATE Given a feasible configuration $\mathcal{S}$, calculate $\boldsymbol{u}(\mathcal{S})$.
\REPEAT
\STATE Set $\mathtt{updated}\gets\mathrm{false}$.
\STATE \emph{Pair construction stage:}
\FOR{each player $\ell\in\mathcal{L}$}
\FOR{each state $s_\ell'\in(\mathcal{K}\times\mathcal{N})\setminus\{s_\ell\}$}
\STATE Construct $\mathcal{S}'$ based on \eqref{construction} and calculate $\boldsymbol{u}(\mathcal{S}')$.
\IF{$\mathcal{S}'\succ\mathcal{S}$}
\STATE Set $\mathcal{S}\gets\mathcal{S}'$.
\STATE Set $\mathtt{updated}\gets\mathrm{true}$.
\ENDIF
\ENDFOR
\ENDFOR
\STATE \emph{Pair exchange stage:}
\FOR{each pair of players $(\ell,i)$ with $\ell<i$}
\STATE Construct $\mathcal{S}'$ based on \eqref{exchange} and calculate $\boldsymbol{u}(\mathcal{S}')$.
\IF{$\mathcal{S}'\succ\mathcal{S}$}
\STATE Set $\mathcal{S}\gets\mathcal{S}'$.
\STATE Set $\mathtt{updated}\gets\mathrm{true}$.
\ENDIF
\ENDFOR
\UNTIL{$\mathtt{updated}=\mathrm{false}$}
\RETURN $\mathcal{S}$.
\end{algorithmic}
\end{algorithm}

Based on the preceding preference relation and configuration updates, the game based user scheduling procedure is summarized in Algorithm~\ref{usalg}. For the first AO iteration, Algorithm~\ref{usalg} is initialized by assigning users consecutive time-slot blocks in ascending order of their $x$-coordinates. Specifically, an equal number of slots is initially allocated to each user, with any remaining slots assigned one per user in this order. Each user is served by its associated antenna throughout its block. For subsequent executions, the configuration optimized in the preceding AO iteration serves as the starting point. In each AO iteration, the antenna trajectories are first updated for the current scheduling configuration and then held fixed throughout the scheduling procedure. Each candidate configuration $\mathcal{S}'$ is accepted if $\mathcal{S}'\succ\mathcal{S}$. After each accepted update, subsequent candidates are constructed based on the updated configuration. Each round includes both pair construction and pair exchange. The algorithm terminates when neither stage accepts an update during a complete round, i.e., $\mathtt{updated}=\mathrm{false}$, indicating stability with respect to both update operations for the fixed trajectories. The resulting configuration is used for trajectory design in the next AO iteration.
%%%%%%%%%%%%%%%%%%%%%%%%%%%%%%%%%%%%%%%%%%%%%%%%%
\subsection{Properties Analysis}
The computational complexity, convergence, and equilibrium properties of the proposed game based user scheduling algorithm are analyzed as follows.

\subsubsection{Complexity}
The computational complexity of Algorithm~\ref{usalg} is analyzed in terms of the number of candidate evaluations. Let $C$ denote the number of scheduling rounds. Each round considers at most $L(KN-1)$ pair-construction candidates and $L(L-1)/2$ pair-exchange candidates. Accordingly, the overall complexity can be expressed as $\mathcal{O}\!\left(C\left[L(KN-1)+L(L-1)/2\right]\right)$.

\subsubsection{Convergence}
Every accepted update satisfies $\mathcal{S}'\succ\mathcal{S}$ and strictly improves the common utility according to \eqref{preference}, although the minimum average rate may remain unchanged. Since the antenna trajectories remain fixed throughout the scheduling procedure, each configuration has a fixed utility vector. Moreover, each of the $L$ players has $KN$ possible scheduling states, yielding at most $(KN)^L$ feasible configurations. Therefore, strict utility improvement prevents any previously visited configuration from being accepted again. Since only finitely many updates can be accepted, Algorithm~\ref{usalg} terminates after a finite number of rounds.

\subsubsection{Stability}
Algorithm~\ref{usalg} terminates when neither stage accepts an update during a complete round. If the final configuration $\mathcal{S}$ were not stable, a feasible candidate $\mathcal{S}'\succ\mathcal{S}$ would be accepted, which contradicts termination. Therefore, the final configuration satisfies both Nash stability and exchange stability for the fixed antenna trajectories, without implying global optimality of problem \eqref{problem}.
%%%%%%%%%%%%%%%%%%%%%%%%%%%%%%%%%%%%%%%%%%%%%%%%%
%%%%%%%%%%%%%%%%%%%%%%%%%%%%%%%%%%%%%%%%%%%%%%%%%
\section{Simulation Results}
\begin{table}[t]
\centering
\caption{Simulation Parameters}\vspace{-1mm}
\label{parameter}
\begin{tabular}{lc}
\hline
\textbf{Parameter} & \textbf{Value} \\ \hline
Carrier frequency ($f_c$) & $28$~GHz\\
Noise power ($\sigma^2$) & $-90$~dBm\\ 
Effective refractive index ($n_\mathrm{eff}$) & $1.4$ \\
Service Period Duration ($T$) & $40$~s \\
Number of Time Slots ($L$) & $40$ \\
Time slot duration ($\Delta t$) & $1$~s \\
Waveguide height ($d$) & $3$~m \\
Initial antenna $x$-coordinate ($x_k[0]$) & $0$~m \\
Service region ($D_x \times D_y$) & $40$~m $\times$ $20$~m \\
Maximum antenna movement speed ($v_\mathrm{max}$) & $2$~m/s \\ 
Convergence tolerance ($\epsilon$) & $10^{-3}$ \\
Maximum number of iterations ($I_\mathrm{max}$) & $20$ \\ \hline
\end{tabular}\vspace{-4mm}
\end{table}

In this section, numerical results are presented to evaluate the performance of the proposed scheme. In this simulation, users are randomly distributed on the ground within the service region. The $K$ waveguides are uniformly spaced at $y_k=D_y/2-(k-1/2)D_y/K$. Unless otherwise specified, the simulation parameters are given in Table~\ref{parameter}.

The initial trajectories and scheduling follow the initialization procedures described for Algorithms~\ref{tdalg} and~\ref{usalg}, respectively. To evaluate the gains from the proposed solution, three algorithm benchmarks are considered: i) initial configuration using both initial trajectories and scheduling; ii) trajectory design using SCA with initial scheduling; and iii) user scheduling using the game based algorithm with initial trajectories. Moreover, two system benchmarks are considered: i) fixed antennas at their initial positions; and ii) instantaneous positioning at the scheduled user's horizontal coordinate without movement delay. Both system benchmarks employ the proposed game based user scheduling algorithm.

\begin{figure}[!t]
\centering
\subfigure[User Distribution]{\includegraphics[width=85mm, trim=0 9mm 0 0, clip]{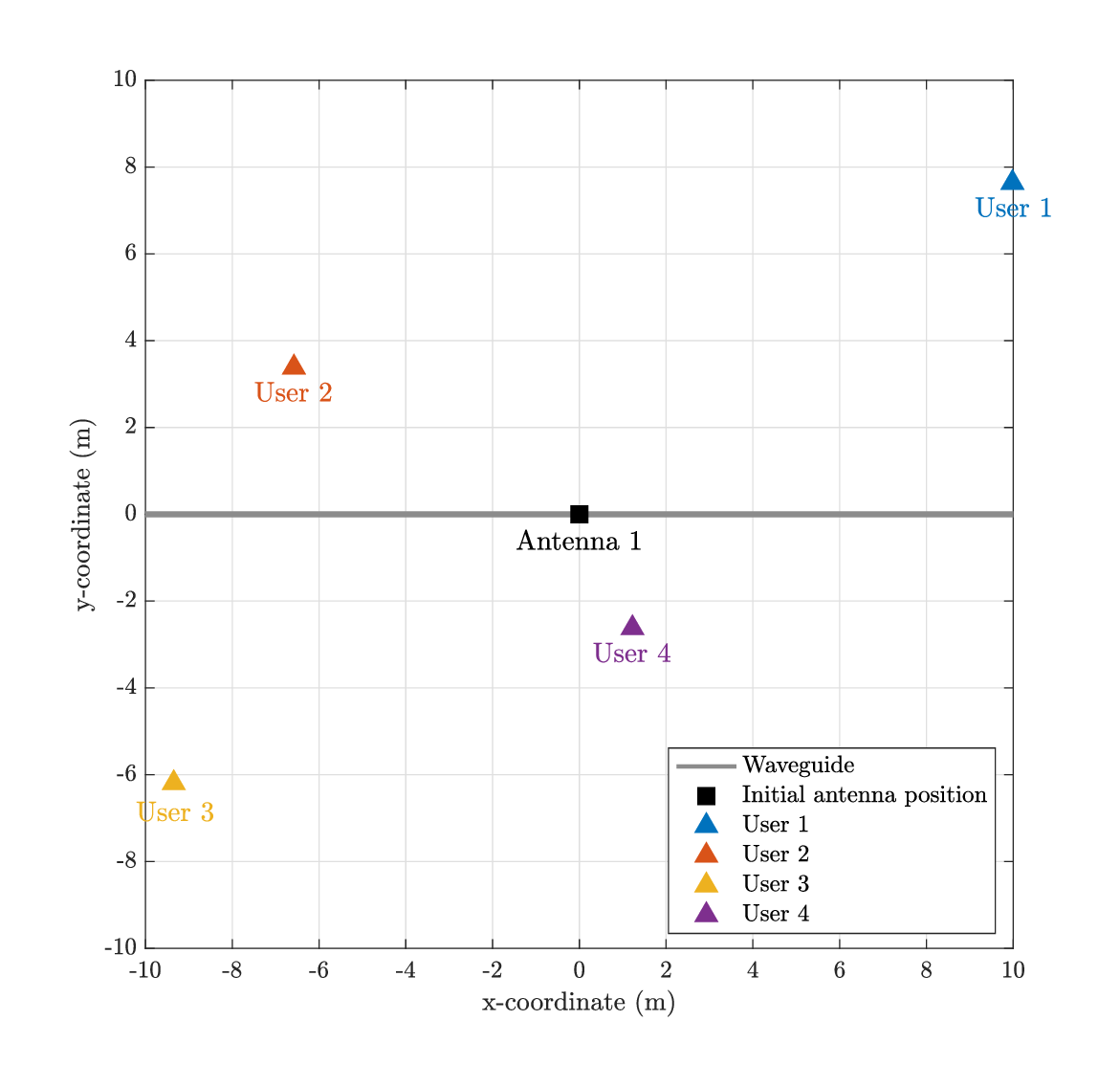}}
\subfigure[Antenna Trajectory]{\includegraphics[width=84mm, trim=0 1mm 0 0, clip]{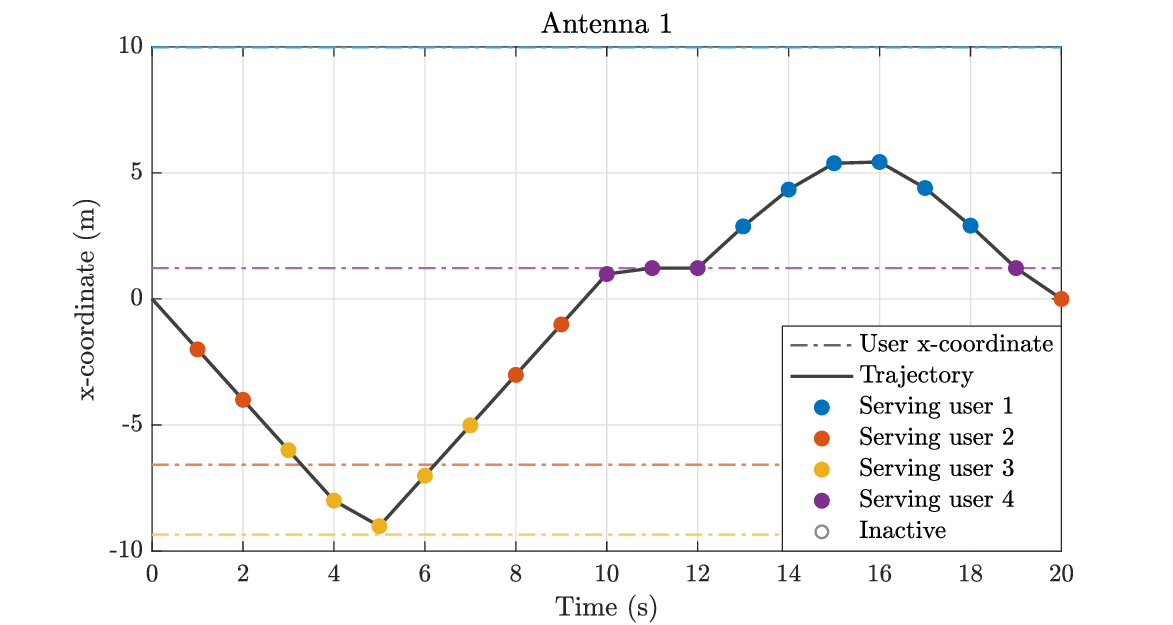}}\vspace{-2mm}
\caption{Illustration of antenna trajectory and user scheduling for a given user deployment, where $D_x=20$~m, $K=1$, $N=4$, $T=20$~s, $L=20$, and $P_t=10$~dBm.}\vspace{-4mm}
\label{result1}
\end{figure}

\begin{figure}[!t]
\centering
\subfigure[User Distribution]{\includegraphics[width=85mm, trim=0 9mm 0 0, clip]{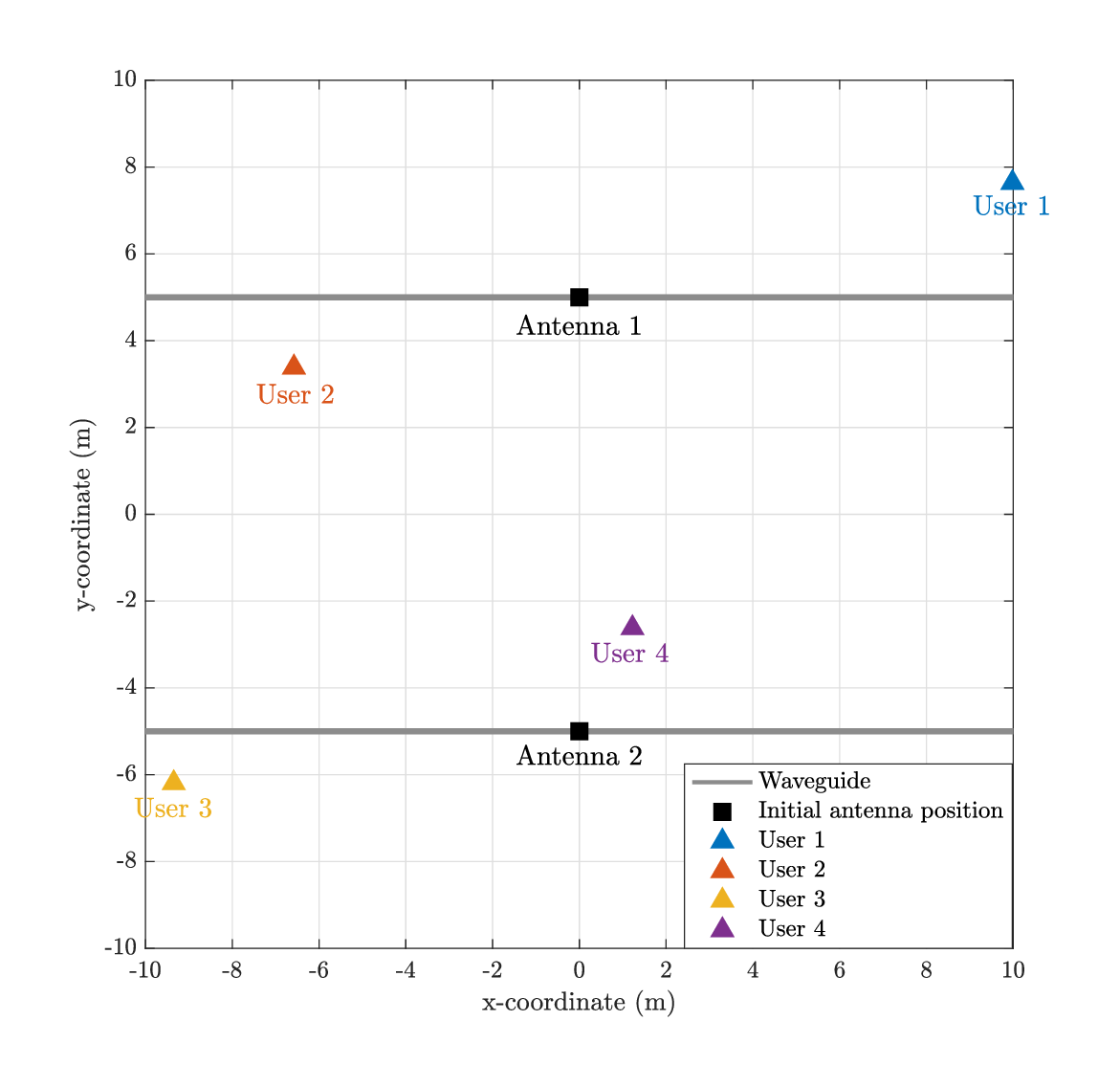}}\vspace{-2mm}
\subfigure[Antenna Trajectory]{\includegraphics[width=84mm, trim=0 17mm 0 0, clip]{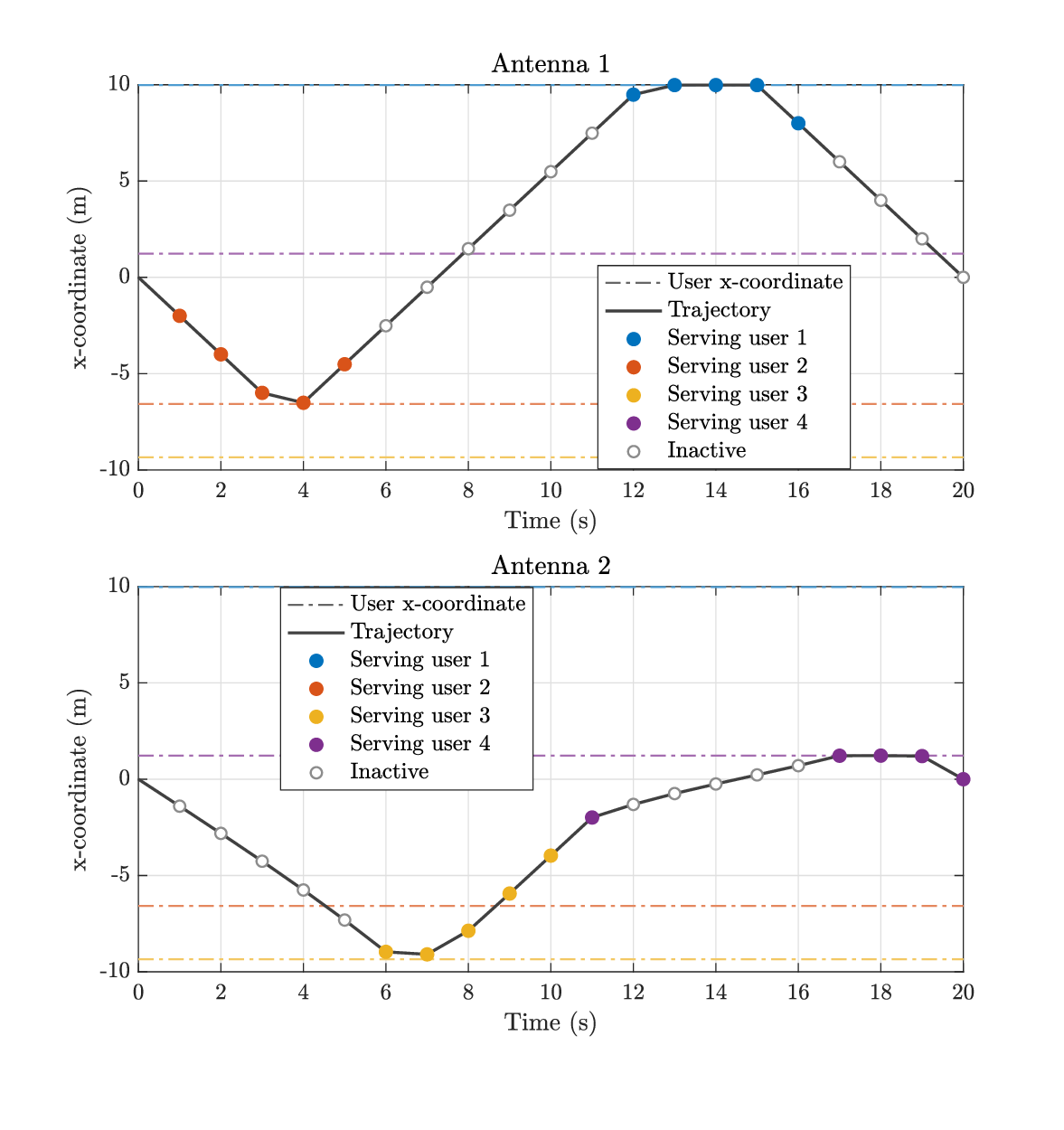}}\vspace{-2mm}
\caption{Illustration of antenna trajectory and user scheduling for a given user deployment, where $D_x=20$~m, $K=2$, $N=4$, $T=20$~s, $L=20$, and $P_t=10$~dBm.}\vspace{-4mm}
\label{result2}
\end{figure}

Figs. \ref{result1} and \ref{result2} illustrate the antenna trajectories and user scheduling for the same user deployment with one and two waveguides, respectively, each equipped with a single antenna. The corresponding minimum average achievable rates are $1.7682$ and $2.1918$ bit/s/Hz. In both cases, user service is coordinated along the trajectories to limit repeated movement between users, which demonstrates the effectiveness of the proposed trajectory design. In \fref{result1}, $6$ time slots are allocated to user $1$ to alleviate its rate bottleneck, while movement and periodicity constraints limit further approach to its optimal position at the waveguide projection. Moreover, user $2$ is served during both the outward and return movements, illustrating how scheduling adapts to the channel conditions along the trajectory. In \fref{result2}, inactive antennas reposition while other users are being served, enabling subsequent transmission from more favorable positions, consistent with Remark~\ref{multiplepin}. This parallel repositioning, together with shorter transverse distances between users and their serving waveguides, contributes to the higher minimum average achievable rate.

\begin{figure}[!t]
\includegraphics[width=85mm]{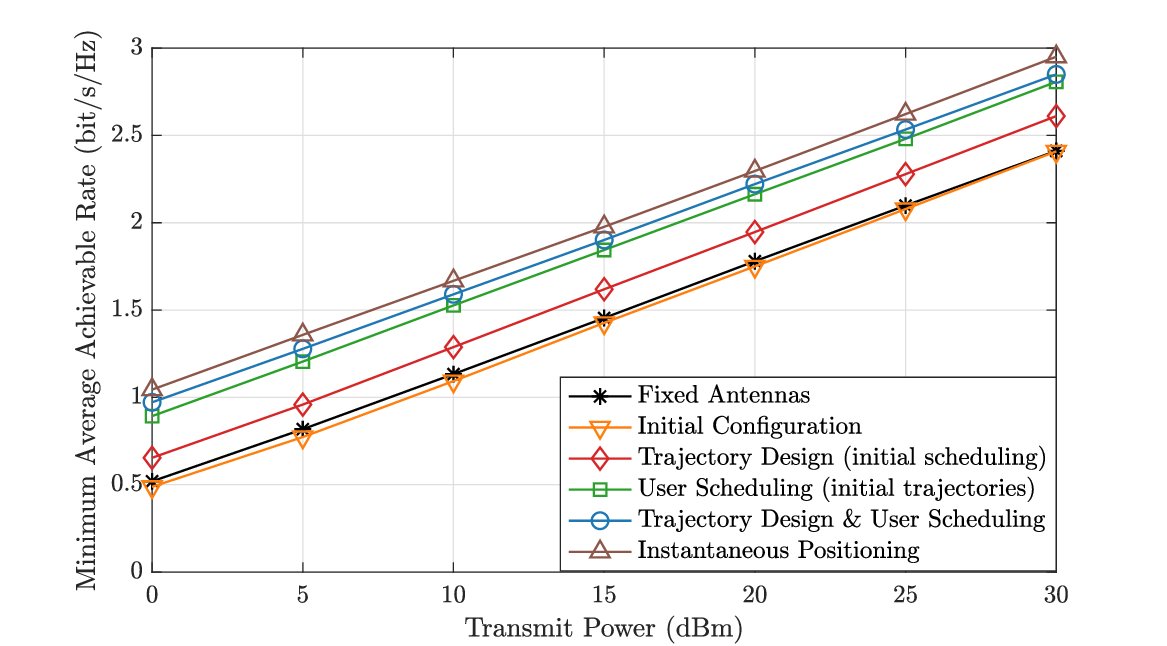}\vspace{-1mm}
\caption{Minimum average achievable rate versus transmit power, where $K=2$, and $N=5$.}\vspace{-4mm}
\label{result3}
\end{figure}

\fref{result3} compares the minimum average achievable rates of the considered schemes versus transmit power. It can be observed that all schemes achieve higher rates as the transmit power increases due to the improved received SNR. Under the considered setting, user scheduling with initial trajectories provides a larger gain than trajectory design with initial scheduling, highlighting the importance of service allocation. The proposed joint design further improves performance and approaches the instantaneous-positioning benchmark, with a small gap remaining due to finite movement speed. Moreover, the fixed-antenna benchmark slightly outperforms the initial configuration, benefiting from optimized user scheduling. This observation indicates that antenna mobility alone does not guarantee a performance improvement when the trajectories and scheduling are not appropriately coordinated.

\begin{figure}[!t]
\includegraphics[width=85mm]{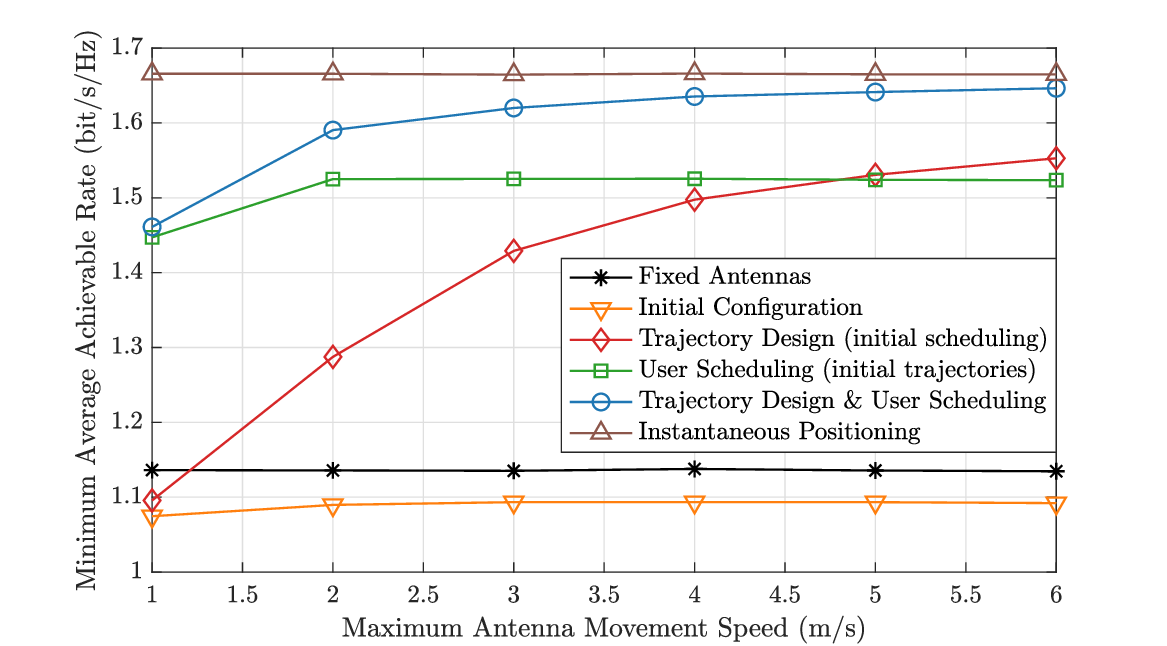}\vspace{-1mm}
\caption{Minimum average achievable rate versus antenna movement speed, where $K=2$, $N=5$, and $P_t=10$~dBm.}\vspace{-4mm}
\label{result4}
\end{figure}

\fref{result4} illustrates the effect of the maximum antenna movement speed on the minimum average achievable rate. As the movement constraints are relaxed, the proposed joint design achieves higher rates and gradually approaches the performance upper bound provided by instantaneous positioning. This trend is consistent with the analytical results, which establish that increasing movement speed reduces the rate loss relative to instantaneous positioning. At low speeds, user scheduling with initial trajectories outperforms trajectory design with initial scheduling, indicating the importance of service allocation when antenna movement is restricted. At higher speeds, trajectory design becomes more effective and eventually surpasses the scheduling benchmark, whose performance saturates as the initial trajectories remain unchanged. Meanwhile, the fixed-antenna and instantaneous-positioning benchmarks remain constant, as their rates are independent of the maximum movement speed.

\begin{figure}[!t]
\includegraphics[width=85mm]{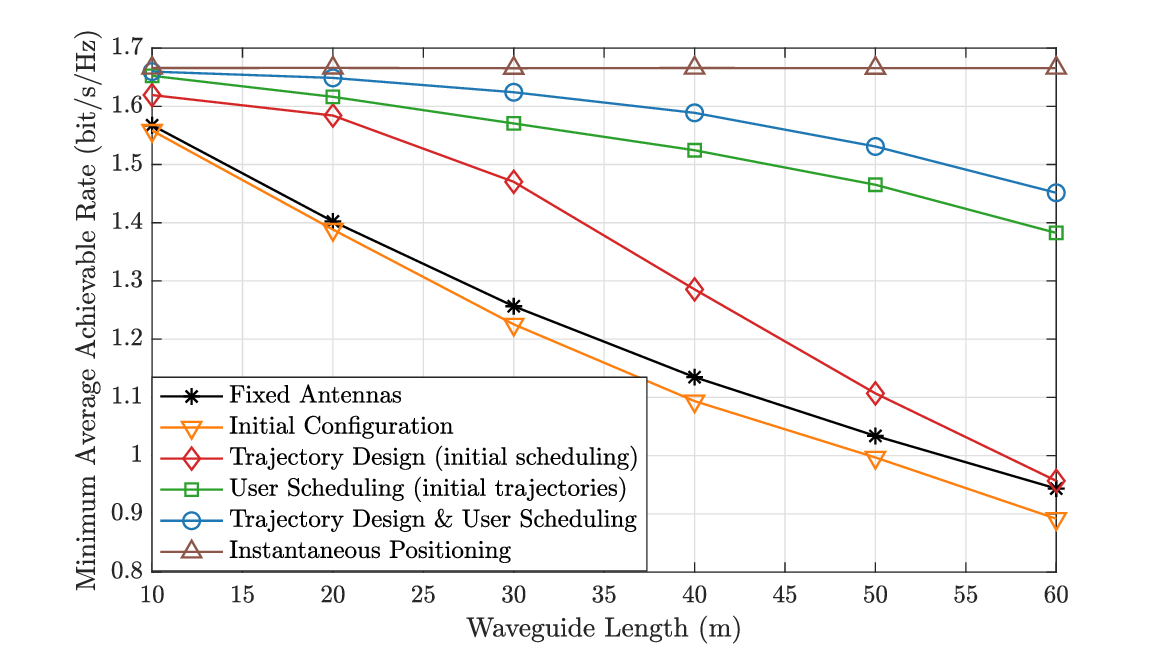}\vspace{-1mm}
\caption{Minimum average achievable rate versus waveguide length, where $K=2$, $N=5$, and $P_t=10$~dBm.}\vspace{-4mm}
\label{result5}
\end{figure}

In \fref{result5}, the effect of waveguide length on the minimum average achievable rate is presented, with the service-region length increasing accordingly. The resulting increase in travel distances makes favorable service positions more difficult to reach within the given service period under finite antenna movement speed. Therefore, the proposed joint design exhibits a growing gap from instantaneous positioning, whose performance remains unchanged due to user-specific antenna positioning. Nevertheless, the rate of the proposed scheme decreases more slowly than those of fixed antennas and the initial configuration, demonstrating the benefits of coordinated trajectory design and user scheduling. Moreover, user scheduling with initial trajectories increasingly outperforms trajectory design with initial scheduling, highlighting the importance of adapting service allocation as the service region expands.

\begin{figure}[!t]
\includegraphics[width=85mm]{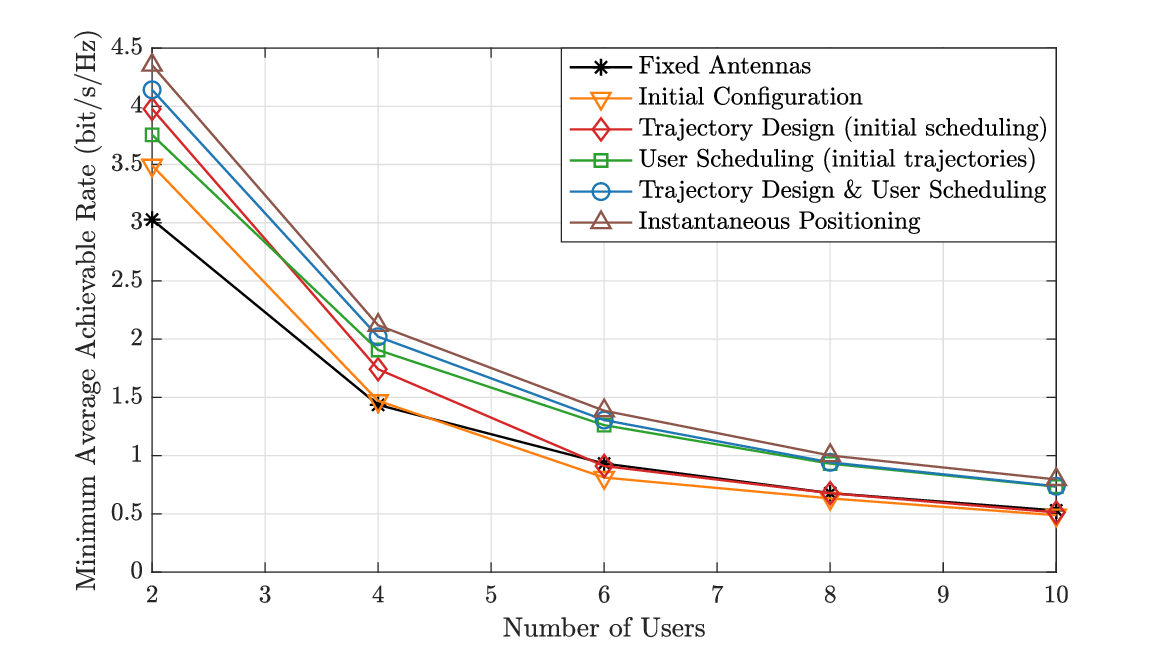}\vspace{-1mm}
\caption{Minimum average achievable rate versus number of users, where $K=2$, and $P_t=10$~dBm.}\vspace{-4mm}
\label{result6}
\end{figure}

The effect of the number of users on the minimum average achievable rate is demonstrated in \fref{result6}. As $N$ increases, more users share the limited service time, reducing the rates of all schemes. The proposed joint design remains close to the instantaneous-positioning benchmark with a decreasing absolute rate gap. Meanwhile, user scheduling with initial trajectories becomes more advantageous than trajectory design with initial scheduling. This is because, with a denser user distribution, scheduling has more opportunities to select nearby users and exploit favorable channel conditions along the antenna trajectories. This flexibility in service allocation highlights the importance of coordinating user service with antenna positions when supporting more users.

\begin{figure}[!t]
\includegraphics[width=85mm]{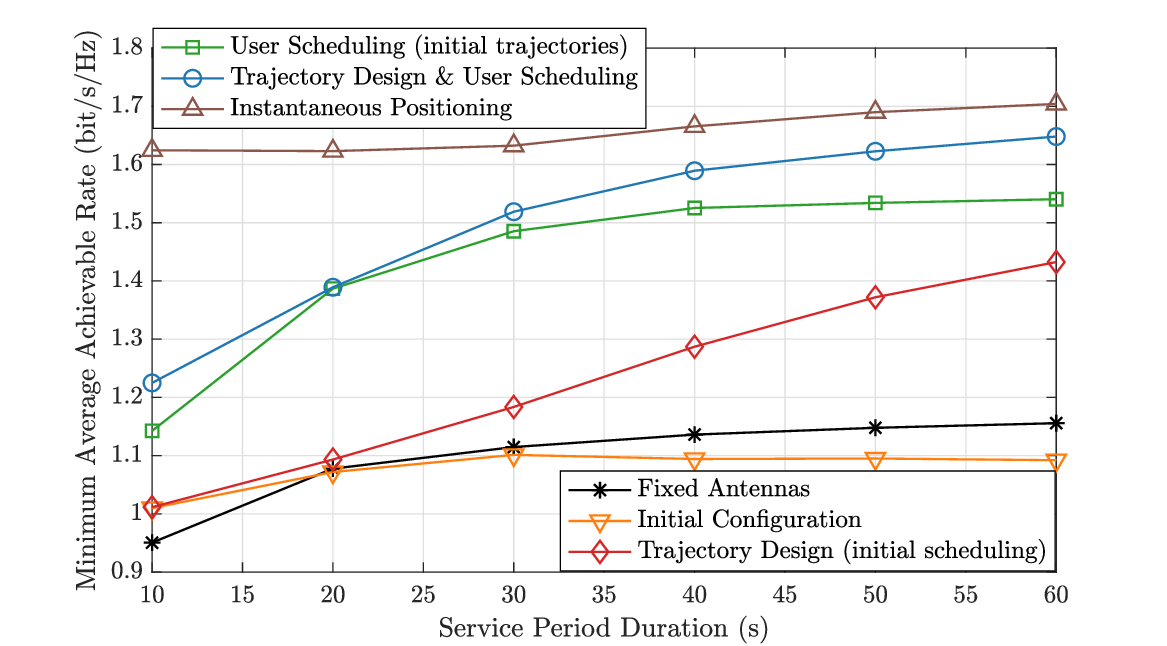}\vspace{-1mm}
\caption{Minimum average achievable rate versus service period duration, where $K=2$, $N=5$, and $P_t=10$~dBm.}\vspace{-4mm}
\label{result7}
\end{figure}

In \fref{result7}, the minimum average achievable rate is evaluated versus the service period duration, with $L$ increasing proportionally to $T$ to maintain $\Delta t=1$ s, ensuring a consistent temporal resolution for trajectory discretization. Specifically, longer service periods provide greater flexibility in antenna movement and service allocation, allowing the proposed joint design to approach the instantaneous-positioning benchmark. At $T=20$ s, user scheduling with initial trajectories achieves nearly the same performance, indicating a limited additional gain from trajectory design at this point. As $T$ further increases, the scheduling benchmark gradually saturates as the initial routes reach their spatial limits, whereas joint design continues to improve by adjusting antenna movement and stationary service. Meanwhile, increasing $L$ enables finer service-time allocation, explaining the slight rate improvements of the fixed-antenna and instantaneous-positioning benchmarks.
%%%%%%%%%%%%%%%%%%%%%%%%%%%%%%%%%%%%%%%%%%%%%%%%%
%%%%%%%%%%%%%%%%%%%%%%%%%%%%%%%%%%%%%%%%%%%%%%%%%
\section{Conclusions}
In this paper, a trajectory based transmission scheme was proposed for pinching-antenna systems, incorporating finite antenna movement speed and transmission during both movement and stationary operation. For a single-waveguide scenario, the expected average rate was derived, and the performance gaps relative to the instantaneous-positioning and fixed-antenna benchmarks were characterized with corresponding bounds. Considering user fairness, a minimum average rate maximization problem was formulated through joint trajectory design and user scheduling. An AO framework combining SCA based trajectory design and game based user scheduling was developed to address this problem. Numerical results demonstrated that the proposed scheme can approach the instantaneous-positioning benchmark under finite antenna movement speed, with joint trajectory design and user scheduling improving the minimum average rate.
%%%%%%%%%%%%%%%%%%%%%%%%%%%%%%%%%%%%%%%%%%%%%%%%%
%%%%%%%%%%%%%%%%%%%%%%%%%%%%%%%%%%%%%%%%%%%%%%%%%
\section*{Appendix~A: Proof of Proposition~\ref{expectrate}}
Under the considered trajectory strategy, each service interval is traversed twice. For user $n$ with $x_n=x_{(i)}$, the left and right service extents are defined as $a_n^\mathrm{L}=x_n-b_{i-1}$ and $a_n^\mathrm{R}=b_i-x_n$, respectively. Thus, the movement distance associated with user $n$ is $S_n=2(a_n^\mathrm{L}+a_n^\mathrm{R})$. The stationary transmission duration allocated to each user is $\tau_\mathrm{s}=\frac{1}{N}(T-S/v_\mathrm{max})$, where $S=\sum_{n=1}^N S_n$ is the total antenna movement distance. Since $S\leq2D_x$, the condition $v_\mathrm{max}T\geq2D_x$ ensures that $\tau_\mathrm{s}\geq0$.

For $N\geq2$, the spacing $G=x_{(i+1)}-x_{(i)}$ between adjacent ordered user coordinates satisfies $\Pr(G>s)=(1-s/D_x)^N$ for $0\leq s\leq D_x$. Each user occupies any given rank in the ordered coordinates with probability $1/N$. Unless user $n$ is the leftmost user, its left service extent equals half an adjacent spacing. For the leftmost user, a positive left service extent occurs only when $x_{(1)}>0$, in which case it equals $x_{(1)}$. Therefore, for a horizontal distance threshold $z\in[0,D_x/2]$, the tail probability of the left service extent is given by
\begin{align}
\Pr(a_n^\mathrm{L}>z)&=\frac{N-1}{N}\left(1-\frac{2z}{D_x}\right)^N\!+\frac{1}{N}\left(\frac{1}{2}-\frac{z}{D_x}\right)^N\nonumber\\
&=\frac{N-1+2^{-N}}{N}\left(1-\frac{2z}{D_x}\right)^N.
\label{pan}
\end{align}
For $N=1$, only the second term in the first line provides a nonzero contribution. By symmetry about $x=0$, the right service extent $a_n^\mathrm{R}$ has the same tail probability. Therefore,
\begin{equation}\label{expecteds}
\mathbb{E}[S]=4N\!\int_0^\frac{D_x}{2}\!\Pr(a_n^\mathrm{L}>z)\mathrm{d}z=\frac{2D_x(N-1+2^{-N})}{N+1}.
\end{equation}

The service extents and stationary transmission duration depend only on the horizontal user coordinates and are independent of the transverse coordinates. Based on \eqref{barr},
\begin{equation}
\bar{r}_n(x_n\pm z)=\frac{1}{D_y}\!\int_{-\frac{D_y}{2}}^\frac{D_y}{2}\!\log_2\!\left(1+\frac{\rho}{d^2+y_n^2+z^2}\right)\mathrm{d}y_n.
\end{equation}
Since the above expression is independent of $x_n$, averaging the stationary and movement contributions over the horizontal user coordinates gives
\begin{align}
\bar{R}_n^\mathrm{traj}&=\frac{1}{N}\left(1-\frac{\mathbb{E}[S]}{v_\mathrm{max}T}\right)\bar{r}_n(x_n)\\
&\quad+\frac{2}{v_\mathrm{max}T}\mathbb{E}\!\left[\int_0^{a_n^\mathrm{L}}\!\bar{r}_n(x_n\!-\!z)\mathrm{d}z\!+\!\!\int_0^{a_n^\mathrm{R}}\bar{r}_n(x_n\!+\!z)\mathrm{d}z\right].\nonumber
\end{align}
Since the two service extents have the same tail probability, exchanging expectation and integration gives
\begin{align}
&\mathbb{E}\!\left[\int_0^{a_n^\mathrm{L}}\!\bar{r}_n(x_n\!-\!z)\mathrm{d}z+\!\int_0^{a_n^\mathrm{R}}\!\bar{r}_n(x_n\!+\!z)\mathrm{d}z\right]\nonumber\\
&=\int_0^\frac{D_x}{2}\!\Pr(a_n^\mathrm{L}>z)\left[\bar{r}_n(x_n\!-\!z)+\bar{r}_n(x_n\!+\!z)\right]\mathrm{d}z\nonumber\\
&=\int_{-\frac{D_x}{2}}^\frac{D_x}{2}\!\Pr(a_n^\mathrm{L}>|z|)\bar{r}_n(x_n\!+\!z)\mathrm{d}z.
\end{align}
In the last equality, $z$ represents the signed horizontal displacement of the antenna from user $n$, and $|z|$ is the corresponding distance. By substituting \eqref{pan} and \eqref{expecteds} into the expected rate expression, \eqref{expectr} is obtained. The proof is completed.\QEDA
%%%%%%%%%%%%%%%%%%%%%%%%%%%%%%%%%%%%%%%%%%%%%%%%%
%%%%%%%%%%%%%%%%%%%%%%%%%%%%%%%%%%%%%%%%%%%%%%%%%
\section*{Appendix~B: Proof of Proposition~\ref{compinst}}
By subtracting \eqref{expectr} from \eqref{expectrinst}, the expected rate loss can be expressed as follows:
\begin{align}
\Delta_n^\mathrm{inst}&=\frac{2D_x(N-1+2^{-N})}{N(N+1)v_\mathrm{max}T}\bar{r}_n(x_n)\\
&\quad-\frac{2(N-1+2^{-N})}{Nv_\mathrm{max}T}\!\!\int_{-\frac{D_x}{2}}^\frac{D_x}{2}\!\!\left(1-\frac{2|z|}{D_x}\right)^N\!\!\bar{r}_n(x_n\!+\!z)\mathrm{d}z.\nonumber
\end{align}
Since $\int_{-D_x/2}^{D_x/2}(1-2|z|/D_x)^N\mathrm{d}z=D_x/(N+1)$, the above expression can be rewritten as \eqref{gapinst}. Moreover, for $\rho>0$, $\bar{r}_n(x_n)>\bar{r}_n(x_n\!+\!z)$ holds for $z\neq0$. Therefore, $\Delta_n^\mathrm{inst}$ is strictly positive for finite $v_\mathrm{max}T$.

To derive the upper bound, a normalized weight is defined for $-D_x/2\leq z\leq D_x/2$, as follows:
\begin{equation}
p(z)=\frac{N+1}{D_x}\left(1-\frac{2|z|}{D_x}\right)^N,
\end{equation}
which satisfies $\int_{-D_x/2}^{D_x/2}p(z)\mathrm{d}z=1$ and $\int_{-D_x/2}^{D_x/2}z^2p(z)\mathrm{d}z=D_x^2/[2(N+2)(N+3)]$. Based on the definitions of $p(z)$ and $\mu$, the rate loss in \eqref{gapinst} can be transformed as follows:
\begin{align}
\Delta_n^\mathrm{inst}&=\frac{2D_x(N\!-\!1\!+\!2^{-N})}{N(N\!+\!1)v_\mathrm{max}T}\!\!\int_{-\frac{D_x}{2}}^\frac{D_x}{2}\!p(z)[\bar{r}_n(x_n)\!-\!\bar{r}_n(x_n\!+\!z)]\mathrm{d}z\nonumber\\
&=\frac{\mu}{N}\int_{-\frac{D_x}{2}}^\frac{D_x}{2}\!p(z)[\bar{r}_n(x_n)-\bar{r}_n(x_n\!+\!z)]\mathrm{d}z.
\end{align}
Based on \eqref{barr}, the rate loss at displacement $z$, averaged over the transverse user coordinate, is given by
\begin{align}\label{rgap}
&\bar{r}_n(x_n)-\bar{r}_n(x_n\!+\!z)\\
=&\mathbb{E}_{y_n}\!\left[\log_2\!\left(1+\frac{\rho}{d^2+y_n^2}\right)-\log_2\!\left(1+\frac{\rho}{d^2+y_n^2+z^2}\right)\right].\nonumber
\end{align}
For each fixed $y_n$, the expression inside the expectation is concave in $z^2$. Since averaging preserves concavity, $\bar{r}_n(x_n)-\bar{r}_n(x_n\!+\!z)$ is also concave in $z^2$. Accordingly, $F(s)=\bar{r}_n(x_n\!+\!\sqrt{s})$ is convex for $s\geq0$ and satisfies $F(z^2)=\bar{r}_n(x_n\!+\!z)$. Since $p(z)$ is nonnegative and integrates to one, applying Jensen's inequality gives
\begin{align}
\Delta_n^\mathrm{inst}&=\frac{\mu}{N}\!\left[\bar{r}_n(x_n)-\int_{-\frac{D_x}{2}}^\frac{D_x}{2}p(z)F(z^2)\mathrm{d}z\right]\nonumber\\
&\leq\frac{\mu}{N}\!\left[\bar{r}_n(x_n)-F\!\left(\int_{-\frac{D_x}{2}}^\frac{D_x}{2}z^2p(z)\mathrm{d}z\right)\right]\nonumber\\
&=\frac{\mu}{N}\!\left[\bar{r}_n(x_n)\!-\!\bar{r}_n\!\left(\!x_n\!+\!\sqrt{\int_{-\frac{D_x}{2}}^\frac{D_x}{2}z^2p(z)\mathrm{d}z}\right)\right].
\end{align}
By substituting the second moment of $p(z)$, the upper bound in \eqref{gapinstupper} is obtained, which completes the proof.\QEDA
%%%%%%%%%%%%%%%%%%%%%%%%%%%%%%%%%%%%%%%%%%%%%%%%%
%%%%%%%%%%%%%%%%%%%%%%%%%%%%%%%%%%%%%%%%%%%%%%%%%
\section*{Appendix~C: Proof of Proposition~\ref{compfix}}
Based on \eqref{barr}, $\bar{r}_n(x_n\!+\!z)$ depends only on $z^2$ and is independent of the particular value of $x_n$. Therefore, \eqref{expectrfix} can be rewritten as follows:
\begin{equation}
\bar{R}_n^\mathrm{fix}=\frac{1}{ND_x}\int_{-\frac{D_x}{2}}^\frac{D_x}{2}\bar{r}_n(x_n\!+\!z)\mathrm{d}z.
\end{equation}
By subtracting this expression from \eqref{expectr} and using $\int_{-D_x/2}^{D_x/2}(1-2|z|/D_x)^N\mathrm{d}z=D_x/(N+1)$, \eqref{gapfix} can be obtained.

To derive the lower bound, the symmetry of $\bar{r}_n(x_n\!+\!z)$ with respect to $z=0$ is used to express \eqref{gapfix} as
\begin{align}
\Delta_n^\mathrm{fix}&=\frac{2}{ND_x}\int_0^\frac{D_x}{2}[\bar{r}_n(x_n)\!-\!\bar{r}_n(x_n\!+\!z)]\mathrm{d}z\\
&\!-\!\frac{4(N\!-\!1\!+\!2^{-N})}{Nv_\mathrm{max}T}\!\!\int_0^\frac{D_x}{2}\!\!\!\left(\!1\!-\!\frac{2z}{D_x}\!\right)^N\!\![\bar{r}_n(x_n)\!-\!\bar{r}_n(x_n\!+\!z)]\mathrm{d}z.\nonumber
\end{align}
Let $\bar{r}_n'(x_n+z)$ denote the derivative of $\bar{r}_n(x_n\!+\!z)$ with respect to $z$. For the first integral, integration by parts gives
\begin{align}
&\frac{2}{ND_x}\!\int_0^\frac{D_x}{2}\!\![\bar{r}_n(x_n)\!-\!\bar{r}_n(x_n\!+\!z)]\mathrm{d}z\nonumber\\
=&\frac{1}{N}\left[\left(\frac{2z}{D_x}\!-\!1\right)[\bar{r}_n(x_n)\!-\!\bar{r}_n(x_n\!+\!z)]\right]_0^\frac{D_x}{2}\nonumber\\
&+\frac{1}{N}\!\int_0^\frac{D_x}{2}\!\!\left(1\!-\!\frac{2z}{D_x}\right)[-\bar{r}_n'(x_n+z)]\mathrm{d}z\nonumber\\
=&\frac{1}{N}\!\int_0^\frac{D_x}{2}\!\!\left(1\!-\!\frac{2z}{D_x}\right)[-\bar{r}_n'(x_n\!+\!z)]\mathrm{d}z,
\end{align}
where the boundary term vanishes because the rate difference is zero at $z=0$ and $2z/D_x-1=0$ at $z=D_x/2$. Similarly, applying integration by parts to the second integral gives
\begin{align}
&\int_0^\frac{D_x}{2}\!\left(1\!-\!\frac{2z}{D_x}\right)^N\![\bar{r}_n(x_n)\!-\!\bar{r}_n(x_n\!+\!z)]\mathrm{d}z\nonumber\\
=&\left[-\frac{D_x}{2(N\!+\!1)}\!\left(1\!-\!\frac{2z}{D_x}\right)^{N+1}[\bar{r}_n(x_n)\!-\!\bar{r}_n(x_n\!+\!z)]\right]_0^\frac{D_x}{2}\nonumber\\
&+\frac{D_x}{2(N\!+\!1)}\!\int_0^\frac{D_x}{2}\!\left(1\!-\!\frac{2z}{D_x}\right)^{N+1}\![-\bar{r}_n'(x_n\!+\!z)]\mathrm{d}z\nonumber\\
=&\frac{D_x}{2(N\!+\!1)}\!\int_0^\frac{D_x}{2}\!\left(1\!-\!\frac{2z}{D_x}\right)^{N+1}\![-\bar{r}_n'(x_n\!+\!z)]\mathrm{d}z.
\end{align}
The boundary term also vanishes at both endpoints. Based on the definition of $\mu$, the coefficient of this integral satisfies
\begin{equation}
\frac{4(N\!-\!1\!+\!2^{-N})}{Nv_\mathrm{max}T}\frac{D_x}{2(N\!+\!1)}=\frac{2D_x(N\!-\!1\!+\!2^{-N})}{N(N\!+\!1)v_\mathrm{max}T}=\frac{\mu}{N}.
\end{equation}
Combining these results gives
\begin{align}\label{gapfix1}
\Delta_n^\mathrm{fix}&=\frac{1}{N}\!\!\int_0^\frac{D_x}{2}\!\!\!\left[\!\left(\!1\!-\!\frac{2z}{D_x}\!\right)\!-\!\mu\!\left(\!1\!-\!\frac{2z}{D_x}\!\right)^{N+1}\right]\![-\bar{r}_n'(x_n\!+\!z)]\mathrm{d}z.
\end{align}
The weight in square brackets is nonnegative because $0<\mu\leq1$ under the feasibility conditions.

Differentiating \eqref{barr} with respect to $z$ gives
\begin{align}
\bar{r}_n'(x_n\!+\!z)&=\frac{2z}{\ln(2)}\mathbb{E}_{y_n}\!\!\left[\frac{1}{d^2\!+\!y_n^2\!+\!z^2\!+\!\rho}\!-\!\frac{1}{d^2\!+\!y_n^2\!+\!z^2}\right]\nonumber\\
&=-\frac{2\rho z}{\ln(2)}\mathbb{E}_{y_n}\!\!\left[\frac{1}{(d^2\!+\!y_n^2\!+\!z^2)(d^2\!+\!y_n^2\!+\!z^2\!+\!\rho)}\right].
\end{align}
For each fixed $z$, the expression inside the last expectation is convex in $y_n^2$. Applying Jensen's inequality and using
\begin{equation}
\mathbb{E}_{y_n}[y_n^2]=\frac{1}{D_y}\!\int_{-\frac{D_y}{2}}^\frac{D_y}{2}y_n^2\mathrm{d}y_n=\frac{D_y^2}{12}
\end{equation}
gives
\begin{equation}
-\bar{r}_n'(x_n\!+\!z)\geq\frac{2\rho z}{\ln(2)\!\left(d^2\!+\!\frac{D_y^2}{12}\!+\!z^2\right)\left(d^2\!+\!\frac{D_y^2}{12}\!+\!z^2\!+\!\rho\right)}.
\end{equation}
For $0\leq z\leq D_x/2$, replacing $z^2$ in the denominator by $D_x^2/4$ gives the further lower bound
\begin{equation}
-\bar{r}_n'(x_n\!+\!z)\geq\frac{2\rho z}{\ln(2)\!\left(d^2\!+\!\frac{D_y^2}{12}\!+\!\frac{D_x^2}{4}\right)\!\left(d^2\!+\!\frac{D_y^2}{12}\!+\!\frac{D_x^2}{4}\!+\!\rho\right)}.
\end{equation}
Since the weight in \eqref{gapfix1} is nonnegative, substituting this bound preserves the inequality direction. Therefore,
\begin{align}
\Delta_n^\mathrm{fix}&\geq\frac{\rho}{N\ln(2)\!\left(d^2\!+\!\frac{D_y^2}{12}\!+\!\frac{D_x^2}{4}\right)\left(d^2\!+\!\frac{D_y^2}{12}\!+\!\frac{D_x^2}{4}\!+\!\rho\right)}\\
&\quad\times\!\left[\int_0^\frac{D_x}{2}\!\!2z\!\left(\!1\!-\!\frac{2z}{D_x}\!\right)\mathrm{d}z\!-\!\mu\!\!\int_0^\frac{D_x}{2}\!\!2z\left(\!1\!-\!\frac{2z}{D_x}\!\right)^{N+1}\!\!\mathrm{d}z\right].\nonumber
\end{align}
The first integral is evaluated as follows:
\begin{equation}
\int_0^\frac{D_x}{2}\!2z\!\left(1\!-\!\frac{2z}{D_x}\right)\mathrm{d}z=\left[z^2\!-\!\frac{4z^3}{3D_x}\right]_0^\frac{D_x}{2}=\frac{D_x^2}{12}.
\end{equation}
For the second integral, substituting $t=2z/D_x$ gives
\begin{align}
\int_0^\frac{D_x}{2}2z\left(1-\frac{2z}{D_x}\right)^{N+1}\mathrm{d}z&=\frac{D_x^2}{2}\int_0^1t(1-t)^{N+1}\mathrm{d}t\nonumber\\
&=\frac{D_x^2}{2(N+2)(N+3)}.
\end{align}
Combining these results, the term in square brackets becomes
\begin{equation}
\frac{D_x^2}{12}-\frac{\mu D_x^2}{2(N+2)(N+3)}=\frac{D_x^2}{12}\left[1-\frac{6\mu}{(N+2)(N+3)}\right],
\end{equation}
which establishes \eqref{gapfixlower}. Finally, $N\geq1$ and $0<\mu\leq1$ imply
\begin{equation}
1-\frac{6\mu}{(N+2)(N+3)}\geq\frac{1}{2}.
\end{equation}
Since all remaining factors in \eqref{gapfixlower} are positive for $\rho>0$, the expected rate improvement is strictly positive. The proof is completed.\QEDA
%%%%%%%%%%%%%%%%%%%%%%%%%%%%%%%%%%%%%%%%%%%%%%%%%
%%%%%%%%%%%%%%%%%%%%%%%%%%%%%%%%%%%%%%%%%%%%%%%%%
\bibliographystyle{IEEEtran}
\bibliography{KaidisBib}

@STRING{IEEE_J_COML       	= "{IEEE} Commun. Lett."}

@STRING{IEEE_J_WCOML       	= "{IEEE} Wireless Commun. Lett."}

@STRING{IEEE_J_COM        	= "{IEEE} Trans. Commun."}

@STRING{IEEE_J_WCOM       	= "{IEEE} Trans. Wireless Commun."}

@STRING{IEEE_M_COM        	= "{IEEE} Commun. Mag."}

@STRING{IEEE_WM_COM        	= "{IEEE} Wireless Commun."}

@STRING{IEEE_J_CST         	= "{IEEE} Commun. Surveys Tutorials"}

@ARTICLE{xu2026loss,
  author={Xu, Yanqing and Ding, Zhiguo and Dobre, Octavia A. and Chang, Tsung-Hui},
  journal=IEEE_J_WCOM, 
  title={Pinching-Antenna System Design With {LoS} Blockage: Does In-Waveguide Attenuation Matter?}, 
  year={2026},
  volume={25},
  number={},
  pages={17322-17337},
  doi={10.1109/TWC.2026.3693541}}

@article{kaidi2026fl,
  title={Age-of-Information Aware Federated Learning with Finite Speed Pinching Antenna},
  author={Wang, Kaidi and So, Daniel K. C. and Ding, Zhiguo},
  journal={arXiv preprint arXiv:2607.23595},
  year={2026}}

@article{chen2026pinching,
  title={Pinching-Antenna Enabled Multicell Wireless Systems},
  author={Chen, Yunshu and Xue, Qing and Hua, Meng and Zhou, Bingpeng and Ma, Shaodan},
  journal={arXiv preprint arXiv:2606.12888},
  year={2026}}

@article{mao2026outage,
  title={Outage-Constrained Environment Division Multiple Access ({EDMA}) for Pinching-Antenna Systems},
  author={Mao, Weihao and Lu, Yang and Xu, Yanqing and Ding, Zhiguo},
  journal={arXiv preprint arXiv:2607.18738},
  year={2026}}

@ARTICLE{gan2026pass,
  author={Gan, Deqiao and Xu, Xiaoxia and Zuo, Jiakuo and Ge, Xiaohu and Liu, Yuanwei},
  journal=IEEE_J_COM, 
  title={Joint Beamforming for {NOMA} Assisted Pinching Antenna Systems ({PASS})}, 
  year={2026},
  volume={74},
  number={},
  pages={2450-2465},
  doi={10.1109/TCOMM.2025.3646822}}

@ARTICLE{zhang2026pin,
  author={Zhang, Runxin and Shao, Yulin and Liu, Yuanwei},
  journal=IEEE_J_WCOM, 
  title={Directional Pinching-Antenna Systems}, 
  year={2026},
  volume={25},
  number={},
  pages={22749-22764},
  doi={10.1109/TWC.2026.3728803}}

@article{kaidi2026pin3,
  title={Leaky-Coaxial Pinching-Antenna System with Adjustable Slot Apertures},
  author={Wang, Kaidi and So, Daniel K. C. and Ding, Zhiguo and Karagiannidis, George K},
  journal={arXiv preprint arXiv:2604.23246},
  year={2026}}

@article{xu2026environment,
  title={Environment-Aware Network-Level Design of Pinching-Antenna Systems--Part {II}: Geometry-Aware Case},
  author={Xu, Yanqing and Ding, Zhiguo and Zhang, Xiu Yin and Duong, Trung Q and Chang, Tsung-Hui},
  journal={arXiv preprint arXiv:2602.17032},
  year={2026}}

@ARTICLE{liu2026pin,
  author={Liu, Peng and Fei, Zesong and Hua, Meng and Chen, Guangji and Wang, Xinyi and Liu, Ruiqi},
  journal=IEEE_J_WCOM, 
  title={Wireless Powered {MEC} Systems via Discrete Pinching Antennas: {TDMA} Versus {NOMA}}, 
  year={2026},
  volume={25},
  number={},
  pages={12034-12049},
  doi={10.1109/TWC.2026.3662288}}

@ARTICLE{lu2026pin,
  author={Lu, Yang and Mao, Weihao and Xie, Xinke and Liao, Shuhai and Xu, Yanqing and Lin, Yushen and Zhang, Ruichen and Ai, Bo and Wang, Xianbin and Ding, Zhiguo},
  journal=IEEE_J_CST, 
  title={A Survey on Pinching-Antenna Systems: Optimization Techniques, Intelligent Designs, and Multifunctional Applications}, 
  year={2026},
  volume={28},
  number={},
  pages={6592-6627},
  doi={10.1109/COMST.2026.3710938}}

@article{kaidi2026roaming,
  title={Pinching-Antenna Systems: From Antenna Placement to Antenna Roaming},
  author={Wang, Kaidi and So, Daniel K. C. and Ding, Zhiguo},
  journal={arXiv preprint arXiv:2608.10136},
  year={2026}}

@ARTICLE{chen2026pin,
  author={Chen, Siyu and Duan, Wei and Gu, Juping and Dang, Shuping and Wen, Miaowen and Ho, Pin-Han},
  journal=IEEE_J_WCOM, 
  title={Pinching-Antenna Systems: Waveguide-Power Loss and Free-Space Path Loss Trade-Off}, 
  year={2026},
  volume={25},
  number={},
  pages={17942-17957},
  doi={10.1109/TWC.2026.3695752}}

@ARTICLE{pan2021ris,
  author={Pan, Cunhua and Ren, Hong and Wang, Kezhi and Kolb, Jonas Florentin and Elkashlan, Maged and Chen, Ming and Di Renzo, Marco and Hao, Yang and Wang, Jiangzhou and Swindlehurst, A. Lee and You, Xiaohu and Hanzo, Lajos},
  journal=IEEE_M_COM, 
  title={Reconfigurable Intelligent Surfaces for {6G} Systems: Principles, Applications, and Research Directions}, 
  year={2021},
  volume={59},
  number={6},
  pages={14-20},
  doi={10.1109/MCOM.001.2001076}}

@ARTICLE{zhu2024movable,
  author={Zhu, Lipeng and Ma, Wenyan and Zhang, Rui},
  journal=IEEE_M_COM, 
  title={Movable Antennas for Wireless Communication: Opportunities and Challenges}, 
  year={2024},
  volume={62},
  number={6},
  pages={114-120},
  doi={10.1109/MCOM.001.2300212}}

@ARTICLE{sun2026pin,
  author={Sun, Mingjun and Ouyang, Chongjun and Wu, Shaochuan and Liu, Yuanwei},
  journal=IEEE_J_WCOM, 
  title={Multiuser Beamforming for Pinching-Antenna Systems: An Element-Wise Optimization Framework}, 
  year={2026},
  volume={25},
  number={},
  pages={6538-6552},
  doi={10.1109/TWC.2025.3625377}}

@ARTICLE{wu2018uav2,
  author={Wu, Qingqing and Zhang, Rui},
  journal=IEEE_J_COM, 
  title={Common Throughput Maximization in {UAV}-Enabled {OFDMA} Systems With Delay Consideration}, 
  year={2018},
  volume={66},
  number={12},
  pages={6614-6627},
  doi={10.1109/TCOMM.2018.2865922}}

@ARTICLE{wu2018uav,
  author={Wu, Qingqing and Zeng, Yong and Zhang, Rui},
  journal=IEEE_J_WCOM, 
  title={Joint Trajectory and Communication Design for Multi-{UAV} Enabled Wireless Networks}, 
  year={2018},
  volume={17},
  number={3},
  pages={2109-2121},
  doi={10.1109/TWC.2017.2789293}}

@ARTICLE{xu2026pin2,
  author={Xu, Yiming and Xu, Dongfang and Yu, Xianghao and Song, Shenghui and Ding, Zhiguo and Schober, Robert},
  journal=IEEE_J_WCOM, 
  title={Joint Radiation Power, Antenna Position, and Beamforming Optimization for Pinching-Antenna Systems With Motion Power Consumption}, 
  year={2026},
  volume={25},
  number={},
  pages={7825-7841},
  doi={10.1109/TWC.2025.3633833}}

@ARTICLE{ding2025analytical,
  author={Ding, Zhiguo and Poor, H. Vincent},
  journal=IEEE_J_WCOM, 
  title={Analytical Optimization for Antenna Placement in Pinching-Antenna Systems}, 
  year={2026},
  volume={25},
  number={},
  pages={21943-21957},
  doi={10.1109/TWC.2026.3722185}}

@ARTICLE{xu2025generalized,
  author={Xu, Yanqing and Cui, Jingjing and Zhu, Yongxu and Ding, Zhiguo and Chang, Tsung-Hui and Schober, Robert and Wong, Vincent W. S. and Dobre, Octavia A. and Karagiannidis, George K. and Poor, H. Vincent and You, Xiaohu},
  journal=IEEE_J_CST, 
  title={Generalized Pinching-Antenna Systems: A Tutorial on Principles, Design Strategies, and Future Directions}, 
  year={2026},
  volume={28},
  number={},
  pages={5872-5908},
  doi={10.1109/COMST.2026.3674222}}

@ARTICLE{kaidi2025pin2,
  author={Wang, Kaidi and Ding, Zhiguo and Karagiannidis, George K.},
  journal=IEEE_J_WCOM, 
  title={Antenna Activation and Resource Allocation in Multi-Waveguide Pinching-Antenna Systems}, 
  year={2026},
  volume={25},
  number={},
  pages={4070-4082},
  doi={10.1109/TWC.2025.3608068}}

@article{liu2025pinching,
  author={Liu, Yuanwei and Wang, Zhaolin and Mu, Xidong and Ouyang, Chongjun and Xu, Xiaoxia and Ding, Zhiguo},
  journal=IEEE_M_COM, 
  title={Pinching-Antenna Systems: Architecture Designs, Opportunities, and Outlook}, 
  year={2026},
  volume={64},
  number={1},
  pages={190-196},
  doi={10.1109/MCOM.001.2500037}}

@article{yang2025pinching,
  author={Yang, Zheng and Wang, Ning and Sun, Yanshi and Ding, Zhiguo and Schober, Robert and Karagiannidis, George K. and Wong, Vincent W.S. and Dobre, Octavia A.},
  journal=IEEE_WM_COM,
  title={Pinching Antennas: Principles, Applications and Challenges}, 
  year={2026},
  volume={33},
  number={2},
  pages={175-184},
  doi={10.1109/MWC.2025.3607867}}

@ARTICLE{xie2025pin,
  author={Xie, Ximing and Fang, Fang and Ding, Zhiguo and Wang, Xianbin},
  journal=IEEE_J_COML, 
  title={A Low-Complexity Placement Design of Pinching-Antenna Systems}, 
  year={2025},
  volume={29},
  number={8},
  pages={1784-1788},
  doi={10.1109/LCOMM.2025.3574633}}

@ARTICLE{kaidi2025pin,
  author={Wang, Kaidi and Ding, Zhiguo and Schober, Robert},
  journal=IEEE_J_WCOML, 
  title={Antenna Activation for {NOMA} Assisted Pinching-Antenna Systems}, 
  year={2025},
  volume={14},
  number={5},
  pages={1526-1530},
  doi={10.1109/LWC.2025.3548280}}

@article{wu2024fluid,
  author={Wu, Tuo and Zhi, Kangda and Yao, Junteng and Lai, Xiazhi and Zheng, Jianchao and Niu, Hong and Elkashlan, Maged and Wong, Kai-Kit and Chae, Chan-Byoung and Ding, Zhiguo and Karagiannidis, George K. and Debbah, Mérouane and Yuen, Chau},
  journal=IEEE_WM_COM, 
  title={Fluid Antenna Systems Enabling {6G}: Principles, Applications, and Research Directions}, 
  year={2026},
  volume={33},
  number={4},
  pages={100-108},
  doi={10.1109/MWC.2025.3629597}}

@article{suzuki2022pinching,
  title={Pinching Antenna-Using a Dielectric Waveguide as an Antenna},
  author={Fukuda, Atsushi and Yamamoto, Hiroto and  Okazaki, Hiroshi and Suzuki, Yasunori and Kawai, Kunihiro},
  journal={NTT DOCOMO Technical J.},
  volume={23},
  number={3},
  pages={5--12},
  year={2022},
  month = {Jan.}}

@ARTICLE{ma2024movable,
  author={Ma, Wenyan and Zhu, Lipeng and Zhang, Rui},
  journal=IEEE_J_WCOM, 
  title={{MIMO} Capacity Characterization for Movable Antenna Systems}, 
  year={2024},
  volume={23},
  number={4},
  pages={3392-3407},
  month={Apr.},
  doi={10.1109/TWC.2023.3307696}}

@ARTICLE{wu2019irs,
  author={Wu, Qingqing and Zhang, Rui},
  journal=IEEE_M_COM, 
  title={Towards Smart and Reconfigurable Environment: Intelligent Reflecting Surface Aided Wireless Network}, 
  year={2020},
  volume={58},
  number={1},
  pages={106-112},
  month={Jan.},
  doi={10.1109/MCOM.001.1900107}}

@article{wong2020fluid,
  title={Fluid antenna systems},
  author={Wong, Kai-Kit and Shojaeifard, Arman and Tong, Kin-Fai and Zhang, Yangyang},
  journal=IEEE_J_WCOM,
  year={2021},
  volume={20},
  number={3},
  pages={1950-1962},
  doi={10.1109/TWC.2020.3037595}
}

@ARTICLE{ding2024pin,
  author={Ding, Zhiguo and Schober, Robert and Poor, H. Vincent},
  journal=IEEE_J_COM, 
  title={Flexible-Antenna Systems: A Pinching-Antenna Perspective}, 
  year={2025},
  volume={73},
  number={10},
  pages={9236-9253},
  doi={10.1109/TCOMM.2025.3555866}}

@book{han2012game,
	Author = {Han, Zhu},
	Publisher = {Cambridge University Press},
	Title = {Game theory in wireless and communication networks: theory, models, and applications},
	Year = {2012}}
%%%%%%%%%%%%%%%%%%%%%%%%%%%%%%%%%%%%%%%%%%%%%%%%%
%%%%%%%%%%%%%%%%%%%%%%%%%%%%%%%%%%%%%%%%%%%%%%%%%
\end{document}